\documentclass[useAMS,usenatbib]{mn2e}

\usepackage{aecompl}
\usepackage[toc,page]{appendix}

\usepackage{graphicx}
\usepackage{amssymb}
\usepackage{amsmath}
\usepackage{aas_macros}
\usepackage{deluxetable}
\usepackage{lscape}
\usepackage{rotating}
\usepackage[symbol*]{footmisc}
\usepackage{gensymb}
\usepackage{tikz}



\title[The role of sub-filaments in fragmentation]{The hierarchical fragmentation of filaments and the role of sub-filaments}
\author[S. D. Clarke et al.]{S. D. Clarke$^{1}$\thanks{E-mail: clarke@ph1.uni-koeln.de }, G. M. Williams$^{2}$ and S. Walch$^{1,3}$. \\$^{1}$I. Physikalisches Institut, Universit{\"a}t zu K{\"o}ln, Z{\"u}lpicher Str. 77, D-50937 K{\"o}ln, Germany \\$^{2}$Centre for Astrophysics Research, School of Physics, Astronomy and Mathematics, University of Hertfordshire,\\ College Lane, Hatfield, Al10 9AB, UK \\$^{3}$Cologne Centre for Data and Simulation Science, University of Cologne, Cologne, Germany \thanks{www.cds.uni-koeln.de}}

\newcommand{\Su}{_{_{\odot}}}
\begin{document}

\date{}

\pagerange{\pageref{firstpage}--\pageref{lastpage}} \pubyear{2002}

\maketitle

\label{firstpage}

\begin{abstract}
  
Recent observations have revealed the presence of small fibres or sub-filaments within larger filaments. We present a numerical fragmentation study of fibrous filaments investigating the link between cores and sub-filaments using hydrodynamical simulations performed with the moving-mesh code \textsc{Arepo}. Our study suggests that cores form in two environments: (i) as isolated cores, or small chains of cores, on a single sub-filament, or (ii) as an ensemble of cores located at the junction of sub-filaments. We term these \textit{isolated} and \textit{hub} cores respectively. We show that these core populations are statistically different from each other. Hub cores have a greater mean mass than isolated cores, and the mass distribution of hub cores is significantly wider than isolated cores. This fragmentation is reminiscent of parsec-scale hub-filament systems, showing that the combination of turbulence and gravity leads to similar fragmentation signatures on multiple scales, even within filaments. Moreover, the fact that fragmentation proceeds through sub-filaments suggests that there exists no characteristic fragmentation length-scale between cores. This is in opposition to earlier theoretical works studying fibre-less filaments which suggest a strong tendency towards the formation of quasi-periodically spaced cores, but in better agreement with observations. We also show tentative signs that global collapse of filaments preferentially form cores at both filament ends, which are more massive and dense than other cores.    

\end{abstract}


\section{Introduction}\label{SEC:INTRO}%

Observations using the Herschel Space Observatory show that filaments act as an intermediate step in the star formation process, linking the gas on the molecular cloud scale with the gas in cores \citep{And10, Arz11, Sch12, Arz13, Kon15, Mar16}. The current filament paradigm argues that molecular clouds first form a complex network of filamentary structures which then proceed to fragment into dense cores \citep{And14}. It is therefore imperative to better understand the fragmentation process of filaments if one wishes to understand how mass accumulates in cores.

Theories suggest that isothermal filaments should fragment when their line-mass is close to the critical line-mass given by:
\begin{equation}
\mu_{_{\rm CRIT}} = \frac{2 c_s^2}{G},
\end{equation}
where $c_s$ is the isothermal sound speed and $G$ is the gravitational constant \citep{Ost64,Lar85,InuMiy92}. Models of equilibrium filaments show that there exists a fastest growing wavelength density perturbation, suggesting that a filament should fragment into a chain of quasi-periodically spaced cores \citep{InuMiy92, InuMiy97, FisMar12}. The spacing of these cores is related to the width of the fragmenting filaments, around 4 times the diameter. The non-equilibrium model presented by \citet{Cla16}, which includes the effects of accretion, shows a more complicated dispersion relation linking perturbation wavelength and growth rate. Yet a fastest growing mode remains which produces quasi-periodically spaced cores when realistic initial density perturbations are used. 

\begin{figure*}
\includegraphics[width=0.95\linewidth]{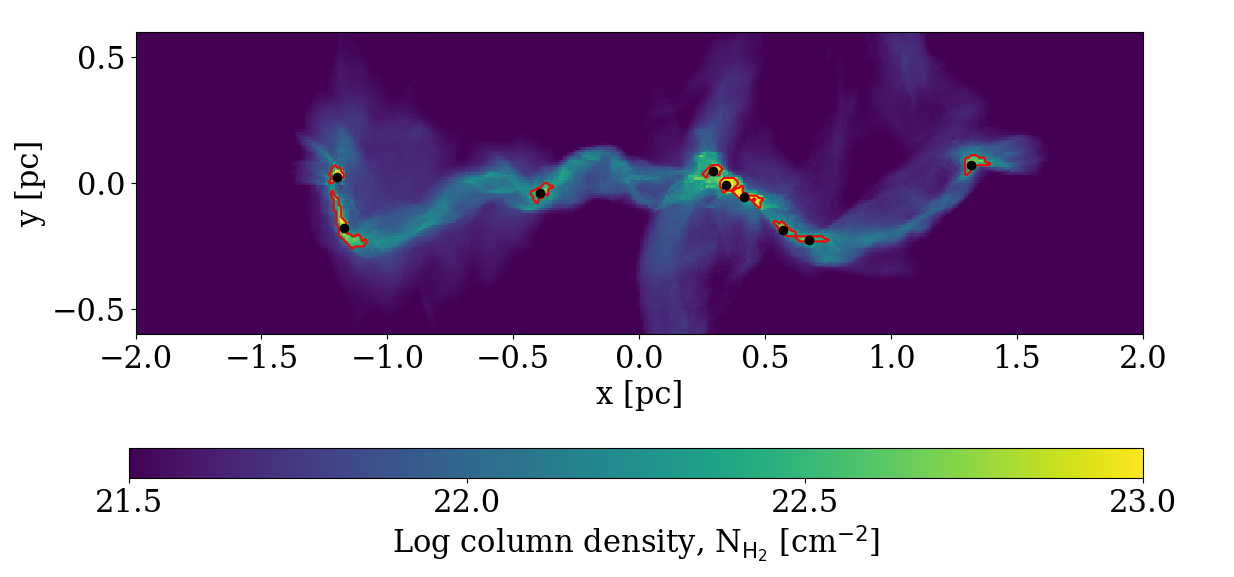}
\caption{An example of the core identification results. The map shows the column density of the filament from \textsc{SIM01}. Overlaid are red contours outlining the boundaries of the cores and black dots showing the column density weighted core location.}
\label{fig::cores}
\end{figure*}  

Studies investigating the substructure of filaments present two challenges to this picture. First, quasi-periodically spaced cores are relatively uncommon \citep{And14}. Second, cores are not the only substructure. Filaments appear to fragment into smaller filaments termed fibres if observed in position-position-velocity (PPV) space or sub-filaments if observed in position-position (PP) or position-position-position (PPP) space \citep{Hac13,TafHac15,Feh16,Dha18,Suri19}. We use this distinction between fibre and sub-filament throughout the paper but note that other authors in the literature differ in their definitions.

Recent simulations agree with observations and show that filaments harbour numerous smaller filaments, though it is currently unclear if these sub-filaments form in situ within the parent filament or form separately and are gathered together into a large scale filament \citep{Smi16,Cla17}. In both scenarios it is the sub-filaments which then go on to fragment into cores. Further, \citet{Cla18} show that the accretion-driven turbulence, which causes the in situ fragmentation of a filament into sub-filaments, leads to the appearance of fibres in synthetic C$^{18}$O observations. While the fibres identified in PPV space are not always directly related to the sub-filaments in PPP space, they propose that as both structures are formed due to the internal turbulence of a filament, a filament which contains fibres also contains sub-filaments. The fact that fibres and sub-filaments are not identical structures is corroborated by cloud-scale simulations presented in \citet{Zam17}. 

In this paper, we present a numerical study of the fragmentation of fibrous filaments focusing on the link between sub-filaments and cores. While the simulations are carried out in 3D we perform this study in 2D, i.e. on column density maps, to more closely resemble observational fragmentation studies. From these simulations we aim to produce a generalised picture of the hierarchical fragmentation of filaments and the role that sub-filaments play in it. In section \ref{SEC:SIM} we detail the simulations used. In section \ref{SEC:TECH} we present the fragmentation analysis techniques used. In section \ref{SEC:IDEN} we describe how we identify and locate cores and how we determine the spines of the sub-filaments and the main filament. In section \ref{SEC:RES} we present the results of the analysis and discuss the link between sub-filaments and core properties, the presence of characteristic fragmentation length-scales, and fragmentation signatures of end-dominated collapse. In section \ref{SEC:CON} we conclude.

\section{Simulations}\label{SEC:SIM}%

The simulations used for this study are the same as those originally presented in \citet{Cla18}, all of which were shown to contain fibres using synthetic C$^{18}$O observations. We summarise the pertinent details here.

The simulations were performed using the moving-mesh code \textsc{Arepo} \citep{Spr10}. The code solves the three-dimensional hydrodynamic equations including self-gravity with time-dependent coupled chemistry and thermodynamics. The current simulations are purely hydrodynamical, the inclusion of a magnetic field is part of a future study as it has been shown to affect fragmentation \citep{Nak93}. 

The computational domain is a slightly flattened box defined by $|x| \leq 3.0$ pc, $|y| \leq 3.0$ pc and $|z| \leq 2.5$ pc. The boundary conditions are periodic with respect to the hydrodynamics but isolated for self-gravity. 

\begin{figure*}
\includegraphics[width=0.95\linewidth]{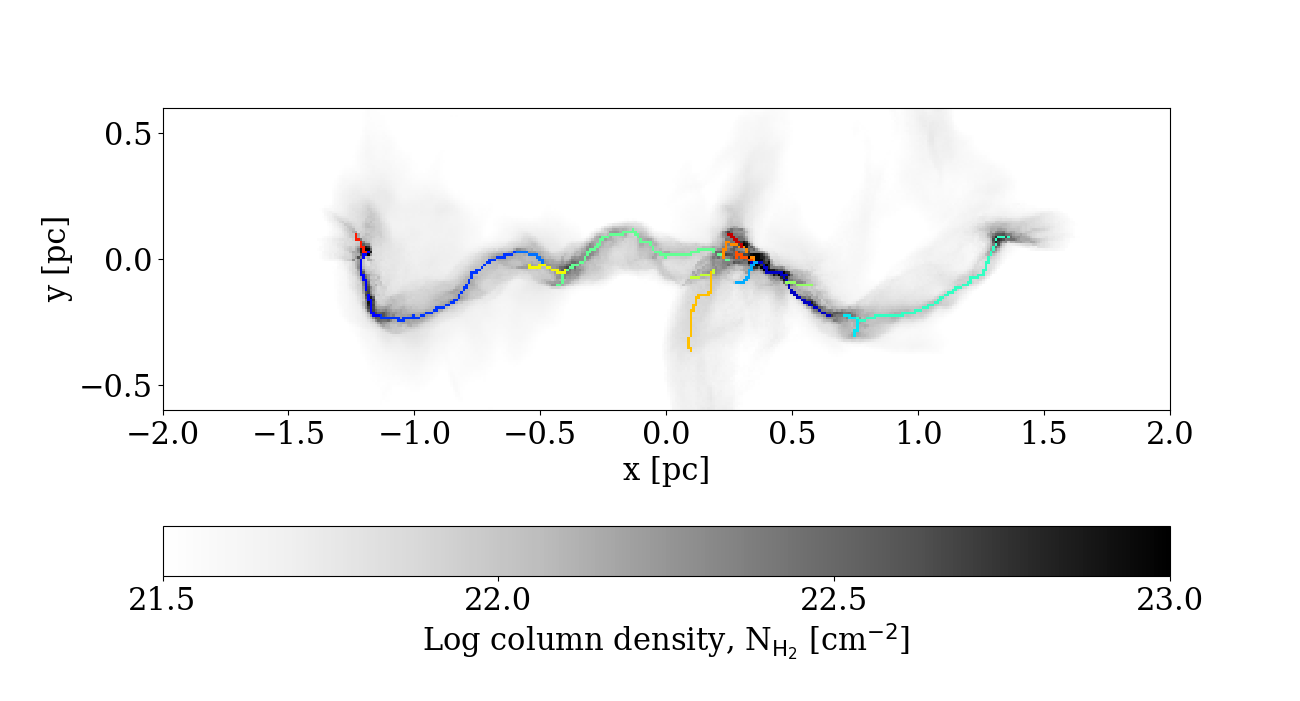}
\caption{A greyscale map showing the log column density of the \textsc{SIM01} filament. Overlaid are the spines of the 16 sub-filaments detected using \textsc{DisPerSe}. }
\label{fig::sub-fils}
\end{figure*}

The initial set-up is an idealised cylindrically symmetric colliding flow. We note that larger-scale influences and non-cylindrical accretion will not be captured by these simulations. Rather the purpose of the simulations is to investigate the basic underlying physics of filament fragmentation without these added complexities. In addition, it allows the simulations to attain the high resolution (see below) necessary to tackle this problem.

The set-up's colliding flow meets at the $z$-axis with an initial velocity of 0.75 km/s and an initial density profile given by $\rho_o / r$, where $\rho_o$ = 15 M$\Su$ pc$^{-3}$ and $r$ is the cylindrical radius. This results in a mass accretion rate from the colliding flow of $\sim$ 70 M$\Su$ Myr$^{-1}$ pc$^{-1}$ towards the $z$-axis. Turbulence is seeded in the inflowing gas with a thermal mix of compressive and solenoidal modes, and a mean velocity dispersion of 1 km/s. It is not driven and allowed to decay. Ten simulations are performed with different turbulent random seeds, labelled \textsc{SIM01} to \textsc{SIM10}. Simulation \textsc{SIM01} is used throughout as an example. 

The set-up quickly forms a dense filament at the $z$-axis. Accretion from the colliding flows drives internal turbulence and the filaments fragment to form numerous cores, see figure \ref{fig::cores}. Mesh cell refinement is used to ensure that the Truelove criterion is always met \citep{Tru97}, i.e. we ensure that the local Jeans mass is resolved by at least 8 cell radii. This requirement leads to a resolution of $\sim 3 \times 10^{-3} - 3 \times 10^{-4}$ pc in the dense gas of the filament, $\rho > 10^{-21} \rm{g \,cm^{-3}}$. Sink particles are inserted at $n \sim 10^7$ cm$^{-3}$ to allow the simulations to proceed after the initial collapse. Here the simulations are stopped when $\sim$ 5 - 10 $\%$ of the mass of the filament is in sinks. At this point the simulations are analysed. This occurs about 0.5 - 0.6 Myr after the start of the simulation and the filaments have reached a mass between $\sim$ 150 M$\Su$and $\sim$ 250 M$\Su$. Before analysis, the \textsc{Arepo} mesh is mapped to a fixed Cartesian grid with resolution of 0.01 pc. We note that our simulations do not include any protostellar feedback effects which may have an effect on core properties and fragmentation; this will be investigated in a forthcoming paper (Clarke et al. in prep.).

\section{FragMent: a tool to study filament fragmentation}\label{SEC:TECH}%

\textsc{FragMent} is an open-source \textsc{Python/C} library\footnote{https://github.com/SeamusClarke/FragMent} presented in \citet{Cla19} which includes a number of fragmentation analysis tools. Here, we use the nearest neighbour separation distribution, the minimum spanning tree edge length distribution, and the two-point correlation function to detect the presence of characteristic fragmentation length-scales. We also make use of the null hypothesis tests included in the package to test the statistical significance of the results. 

A modification has been made to the two-point correlation function included in \textsc{FragMent}. We follow \citet{Gra06} by estimating the error of the two-point correlation function due to Poisson noise with the relation:
\begin{equation}
\sigma_w (r) = \sqrt{\frac{1 + |w(r)|}{DD(r)}},
\end{equation}  
where $w(r)$ is the two-point correlation function evaluated at separation $r$ and $DD(r)$ is the distribution of separation distances of the complete graph constructed from the core locations. This enables us to quantify the significance of any characteristic fragmentation length-scale detected with the two-point correlation function.

Several modifications have been made to the null hypothesis test functions which result in more than an order of magnitude speed up. These modifications are included in the latest version of \textsc{FragMent}.

\begin{figure*}
\includegraphics[width=0.95\linewidth]{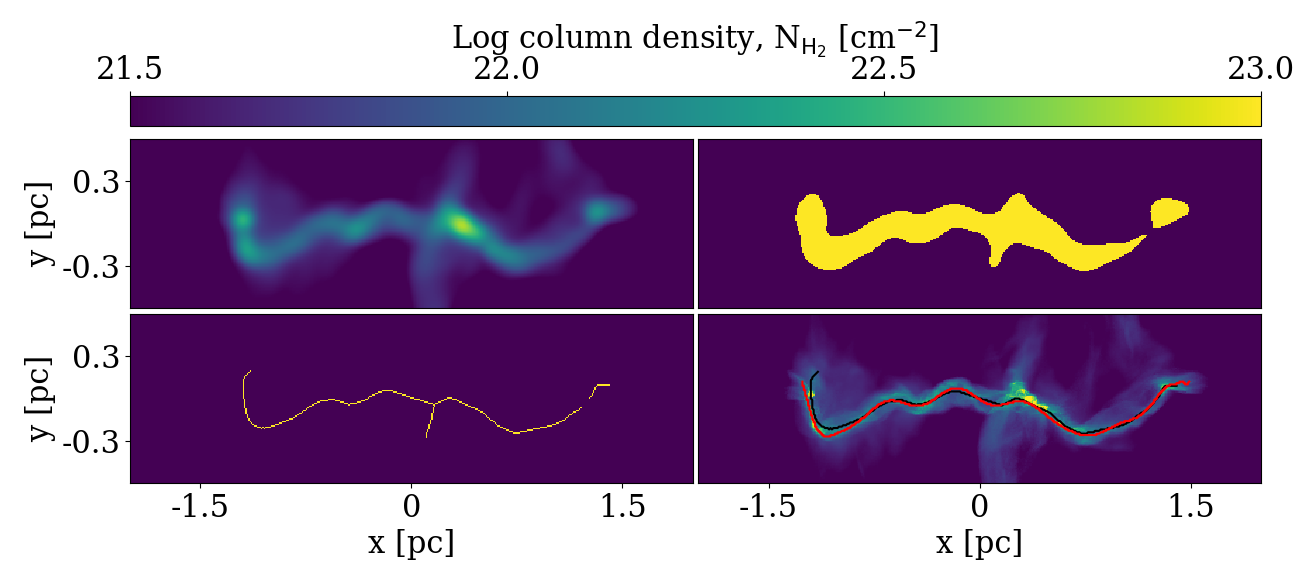}
\caption{A 4 panel image showing the process by which the spine of the main filament is found. (Top left) The column density map is convolved with a Gaussian beam with a standard deviation of 4 pixels. (Top right) A column density threshold cut is applied to produce a binary image. (Bottom left) The binary image of the skeleton produced using the medial axis transform. (Bottom right) The spine of the filament once side-branches have been trimmed and overlaid in white on the unconvolved column density map. A modified spine which passes through high column-density ridges is also shown as a red line.}
\label{fig::Main_spine}
\end{figure*}

\section{Structure identification}\label{SEC:IDEN}%

\subsection{Core identification}\label{SEC:CORE}%

Cores are identified using the dendrogram python package \textsc{astrodendro}\footnote{http://www.dendrograms.org/}. The parameters used for the dendrogram are: \textsc{min$\_$value} = $2 \times 10^{22} \; \rm{cm^{-2}}$, \textsc{min$\_$delta} = $10^{22} \; \rm{cm^{-2}}$, \textsc{min$\_$npix} = 9. Due to the high value of \textsc{min$\_$value} breaking the filament into numerous high column-density regions, the resulting leaves are relatively insensitive to the choice of \textsc{min$\_$delta}. The value of the parameter \textsc{min$\_$npix} leads to a minimum effective diameter of a leaf of 0.03 pc; this helps to ensure that cores are not overly segmented. We identify the leaves of the dendrogram as cores and take their column density weighted centre as the core location. Figure \ref{fig::cores} shows an example of this process, the red contours show the outline of the cores and the black dots the core locations. Over the total of 10 simulations, 116 cores are identified, giving an average of 11.6 cores per filament. The minimum number of cores detected in a filament is 9, and the maximum is 16. 

\subsection{Spine identification}\label{SEC:SPINE}%

As noted in \citet{Cla18}, the simulated filaments contain numerous, smaller sub-filaments. \citet{Cla18} detect these in position-position-position (PPP) space, but they are also apparent in the column density maps of these filaments (PP space). This makes the identification of a spine more complicated. Here we use two different techniques: one to determine the spines of each sub-filament and one to determine the spine of the whole filament. 

\subsubsection{Sub-filaments}\label{SSEC:SUBFIL}%

To find the spines of the sub-filaments seen in column density we used the tool \textsc{DisPerSe} \citep{Sou11}. \textsc{DisPerSe} has been used to identify filaments in numerous observations using column density and line emission maps and has been shown to be sensitive to substructures \citep[e.g.][]{Suri19}. We run \textsc{DisPerSe} using a persistence value of $2 \times 10^{21} \; \rm{cm^{-2}}$, and a threshold of $2 \times 10^{21} \; \rm{cm^{-2}}$. The persistence and threshold are relatively low as we wish to detect sub-structures, not just the prominent main spine. Before analysis we use the inbuilt \textsc{DisPerSe} function to smooth the spines with a smoothing length of 5 pixels. We also use the \textsc{assemble} option to join spines which meet but are at an angle to each other; we use a value of 70 degrees so as to join as many spines as possible together. Finally, we exclude all spines which consist of fewer than 10 pixels as they are mostly artefacts.  

Figure \ref{fig::sub-fils} shows an example of the sub-filaments found using this technique. Here 16 sub-filaments are found. One can see that they relate to column density ridges, and that the high column density region at $x \; \sim \; 0.3$ pc shows a high number of sub-filaments due to its complexity. 

Over all 10 simulations, 148 sub-filaments are found. The minimum number of sub-filaments identified in a simulation is 10 and the maximum is 19. \citet{Cla18} analyse these simulations using synthetic C$^{18}$O observations, though at a slightly earlier time ($\sim 0.1$ Myr earlier), and find that each filament contained on average 22 fibres in PPV space. When identifying sub-filaments in PPP space, they find on average 26 per filament. It is clear that the level of substructure and complexity which is readily identifiable lessens as one uses a reduced amount of information with respect to the underlying 3D density field.

\begin{figure*}
\includegraphics[width=0.95\linewidth]{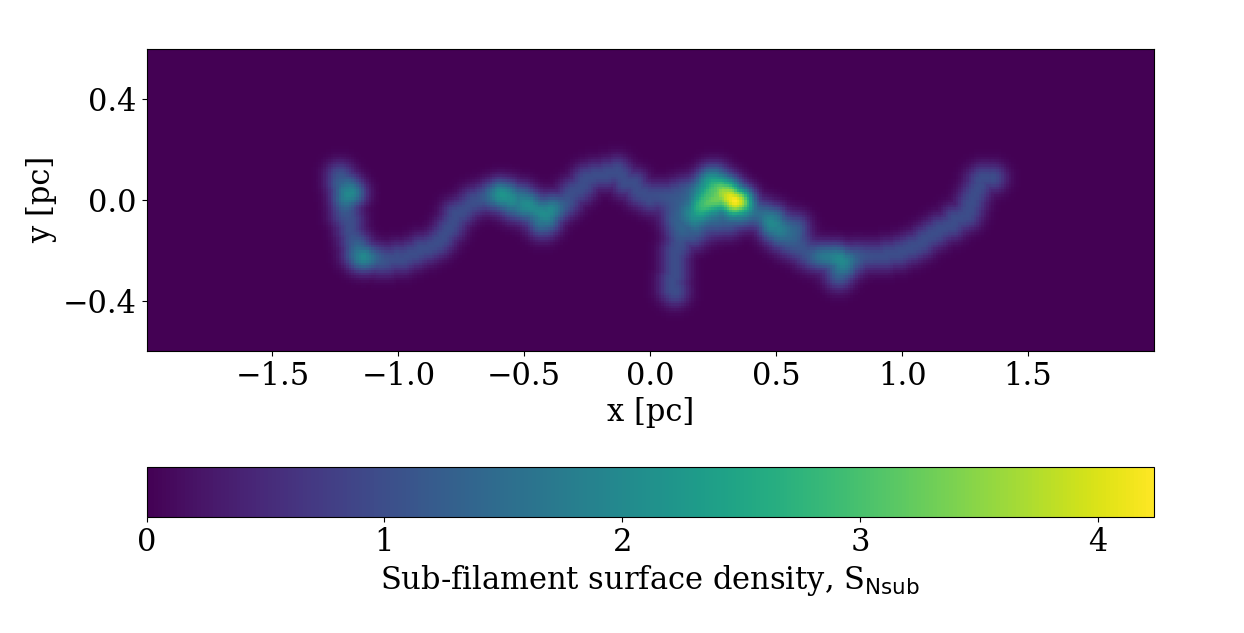}
\caption{A map showing the dimensionless surface density of the number of identified sub-filaments as defined by equation \ref{eq::surface_density}. Its value is close to one over most of the filament but rises as high as $\sim$ 4 in complex regions.}
\label{fig::sub_surface}
\end{figure*}

\subsubsection{Main filament}\label{SSEC:MAINFIL}%

To apply the tools in \textsc{FragMent} one needs the spine of the entire filament, hereafter main filament, rather than the individual spines of the column density sub-filaments picked out by \textsc{DisPerSe}. Here we detail a method similar to that described in \citet{Sch14}. Each step can be seen in figure \ref{fig::Main_spine}.

The first step is to convolve the column density map with a Gaussian beam (figure \ref{fig::Main_spine}, top left panel), this is to dilute the substructure present so that it does not affect the spine. However, if the beam size is too large the resulting spine is often shifted from that which one would identify by eye. We find that a standard deviation of between 3 and 5 pixels is sufficient to hide substructure without distorting the spine. We use a standard deviation of 4 pixels for all 10 simulations.

A column density threshold is applied to the convolved column density map to identify the entirety of the filament (figure \ref{fig::Main_spine}, top right panel). This produces a binary image, i.e. one consisting of 1s and 0s, which shows the filament. If the column density threshold is set too low, features seen in the turbulent accretion flow may be detected. Set too high and the filament is broken into small pieces. A value of $4 \times 10^{21}$ cm$^{-2}$ is used here, for all simulations, as it is a good compromise for these simulated filaments.

To produce the skeleton of this binary image we use the medial axis transformation included in the \textsc{Python} package \textsc{scikit-image} \citep{scikit}. The median axis is defined as the set of points each having more than one closest point on the image boundary. It is therefore able to reduce an image to a one pixel wide skeleton while preserving the general morphology. 

As seen in the bottom left panel of figure \ref{fig::Main_spine}, the result from the medial axis transformation is not the spine of the filament; it contains unwanted side-branches and in some cases there exist small gaps in the skeleton. These side-branches must be removed and the gaps must be filled manually. The result is a continuous, single, one pixel wide skeleton, which traces the spine of the main filament, seen as the black line in the bottom right panel of figure \ref{fig::Main_spine}. However, this process may result in areas where the spine does not pass through clear high column density ridges. Therefore, small modifications are made manually to force this to be the case; this is seen as the red line in the bottom right panel of figure \ref{fig::Main_spine}. This is the main filament spine used.

\section{Results}\label{SEC:RES}%

\subsection{Link between cores and sub-filaments}\label{SSEC:SUBCORE}%

\begin{figure}
\includegraphics[width=0.95\linewidth]{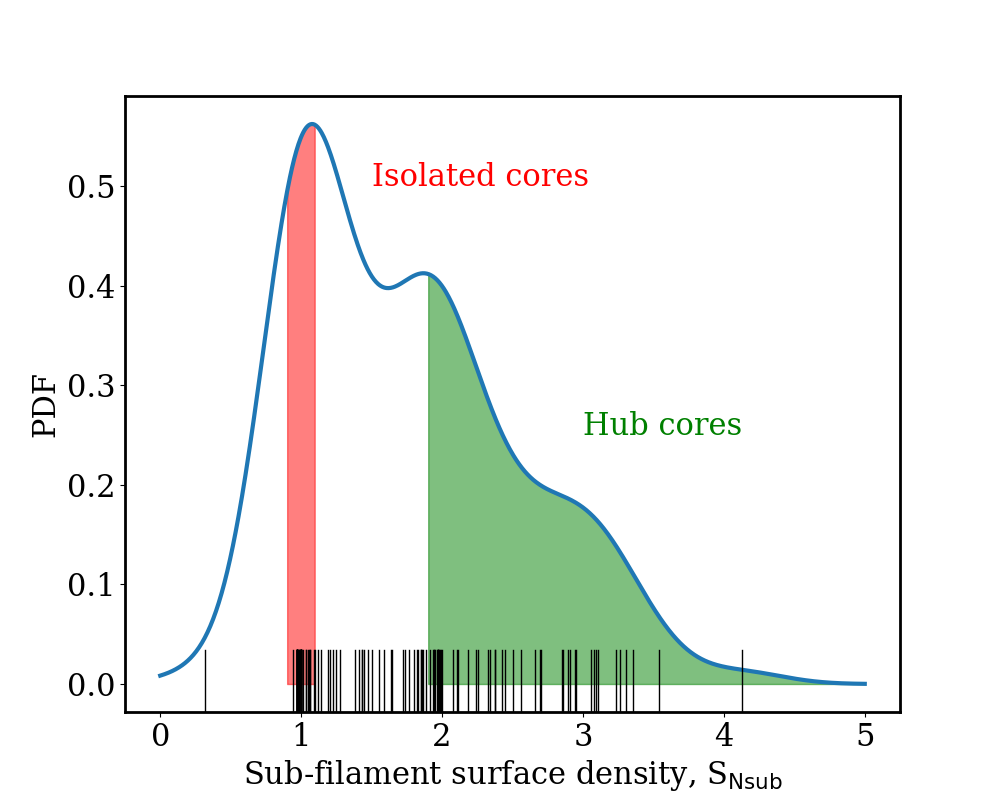}
\caption{A KDE showing the distribution of sub-filament surface density, $S_{\rm Nsub}$, at the 116 core locations across all 10 simulations. The small vertical lines show each individual data point. Cores in the red shaded region, $0.9 \leq S_{\rm Nsub} \leq 1.1$, are termed \textit{isolated} cores and those in the green shaded region, $S_{\rm Nsub} \geq 1.9$, are termed \textit{hub} cores. The distribution is normalised such that the integral is equal to 1. The KDE bandwidth is 0.18.}
\label{fig::SFSDvsCORE}
\end{figure}

It is clear that the cores lie on, or very close to, the spine of the main filaments (see figures \ref{fig::cores} and \ref{fig::Main_spine}). However, it is unclear how the cores are related to the sub-filaments identified in the column-density map. In similar simulations to those analysed here, \citet{Cla17} note that a number of the elongated structures in PPP space overlap and merge to form hub-like structures in which cores form. In observations studying fibres in PPV space, \citet{TafHac15} propose a scenario of filament fragmentation termed `\textit{fray and fragment}' which claims that filaments fragment into numerous fibres which then proceed to fragment into cores independently of each other. This scenario may also occur in the sub-filaments identified here in PP space. Here we attempt to characterise the link between cores and sub-filaments.

We construct a dimensionless measure of the surface density of sub-filaments across the column density map. This is done by considering the minimum distance between each sub-filament and a pixel. The sub-filament surface density $S_{\rm Nsub}$ for pixel $j$ is given by the equation:

\begin{equation}
S_{\rm{Nsub},j} = \sum_i^N e^{-r_{min,ij}^2 / 2 \sigma^2_{\rm Nsub}} \; ;
\label{eq::surface_density}
\end{equation}

where $r_{min,ij}$ is the minimum distance between the pixel co-ordinates, ($x_j$,$y_j$), and one of the spine pixels of sub-filament $i$, $\sigma_{\rm Nsub}$ is the bandwidth of the kernel, and $N$ is the number of sub-filaments. The idea for this quantity is that each sub-filament's contribution to a pixel's surface density is weighted by a Gaussian kernel of the minimum distance between the pixel and the sub-filament. As the number of sub-filaments increases, or the sub-filaments become closer to the pixel, the measure $S_{\rm Nsub}$ increases; it thus acts similar to a surface density. As only the minimum distance to each sub-filament is used, this measure is not affected by how many sub-filament spine pixels are nearby but only how many different sub-filament spine pixels are. The term is dimensionless and the exact value is not important (it may range from 0 to $N$), but the relative values across the map are important. We take $\sigma_{\rm Nsub}$ to be 4 pixels here. The exact value chosen is arbitrary. We take $\sigma_{\rm Nsub} = 4$ which results in a kernel with a full width half maximum of $\sim$ 0.1 pc as our column density map resolution is 0.01 pc.

\begin{table*}
\centering
\begin{tabular}{@{}*7l@{}}
\hline\hline
Core population     & Number of cores & Median & Interquartile range & Mean & Standard deviation & Range      \\ \hline
All cores           & 116             & 3.3    & 3.8                 & 4.7  & 4.4                & 0.6 - 24.9 \\
Isolated cores ($0.9 \leq S_{\rm Nsub} \leq 1.1$)     & 38              & 2.6    & 2.4                 & 2.9  & 1.8                & 0.7 - 8.5  \\
Hub cores ($S_{\rm Nsub} \geq 1.9$)          & 47              & 4.5    & 5.4                 & 6.2  & 5.6                & 0.6 - 24.9 \\ 
Intermediate cores ($1.1 \leq S_{\rm Nsub} \leq 1.9$) & 30 & 2.9 & 3.6 & 4.2 & 3.3 & 0.6 - 12.5 \\ \hline
\end{tabular}
\centering
\caption{A summary of the core mass distribution properties for the different core populations. All masses are in solar masses.}
\label{tab::core_mass}
\end{table*}

Figure \ref{fig::sub_surface} shows the surface density of sub-filaments for \textsc{SIM01}. The majority of the filament shows a value close to 1, denoting a close presence of only one sub-filament. There are bright spots with a surface density of 2, showing areas where two sub-filaments meet. The complex region located at $x \; \sim \; 0.3$ pc shows a surface density as high as 4.

\begin{figure}
\includegraphics[width=0.95\linewidth]{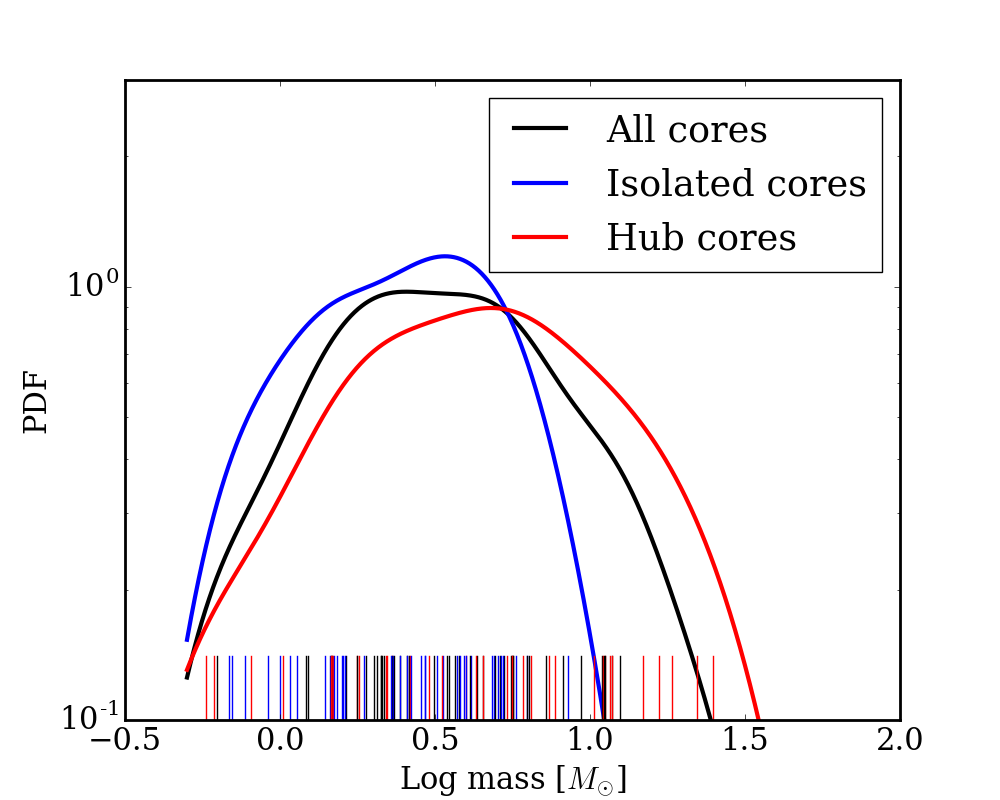}
\caption{A probability density function of the core masses considering all 116 cores from the 10 simulations (black), the isolated core population (blue) and the hub core population (red). The small vertical lines show each individual data point. The distribution is normalised such that the integral is equal to 1. The KDE bandwidth is 0.08 for the all core data set, 0.10 for the isolated core data set, and 0.16 for the hub core data set.}
\label{fig::CMF}
\end{figure}

Figure \ref{fig::SFSDvsCORE} shows a Kernel Density Estimator (KDE) of the surface density of sub-filaments at the core locations determined in section \ref{SEC:CORE}. The bandwidths of all KDEs are calculated using Scott's rule \citep{Scott}. There is a large grouping of core locations where the surface density is close to 1 ($\sim 33\%$ of all cores have surface densities between 0.9 and 1.1), showing that they lie on or very close to a single sub-filament. This is consistent with the idea of the fray and fragment model, that each sub-filament fragments into a core, or cores, independently of each other. We call these cores \textit{isolated cores}. However, there are a significant number of cores where the surface density is 1.9 or greater ($\sim 41\%$ of all cores) showing that there exist hub-sub-filament systems within the main filament, as proposed by \citet{Cla17}. We call these cores \textit{hub cores}. There also exists some cores which lie in between these extremes ($\sim 26\%$ of all cores) which we call \textit{intermediate cores.} These are typically cores on the edge of hubs.

One may ask if the cores formed in individual sub-filaments differ from those formed at the junctions of multiple sub-filaments. We first focus on their mass. Figure \ref{fig::CMF} shows a KDE of the core mass distribution considering all 116 cores. There is no significant difference between the cores found in the 10 simulations. The distribution appears close to a log-normal, though with a flat top. However, we do not expect to fully represent the observed core mass function due to the idealised nature of the setup.  

Figure \ref{fig::CMF} also shows the core mass distributions for the isolated and the hub cores. It is clear that there is a difference in the core mass distribution between these populations; isolated cores tend to lower masses and have a narrower distribution. A two sample Kolmogorov-Smirnov test, a null hypothesis test where the null is that the two samples come from the same underlying distribution, returns a large distance statistic, 0.39, with the corresponding p-value of 0.005. We can thus confidently reject the null hypothesis and assert that the two populations have a significantly different core mass distribution. We also perform a two-sided Mann-Whitney U test \citep{Mann47}. The Mann-Whitney U test is a non-parametric test and its null hypothesis is typically stated as neither distribution having stochastic dominance over the other. This can formally be expressed as the probability of a variable drawn from distribution X having a greater value than a variable drawn from distribution Y being equal to the reverse, i.e. $P(X>Y) = P(Y>X)$. Thus the test is sensitive to differences in location, width and form between the two tested distributions, similar to the KS test. Using the Mann-Whitney U test we are able to reject the null hypothesis at high confidence, a $p$-values of 0.0036, and can therefore state that hub cores have a greater mass than isolated cores on average. Table \ref{tab::core_mass} summaries the properties of these core mass distributions. 

\begin{figure}
\includegraphics[width=0.95\linewidth]{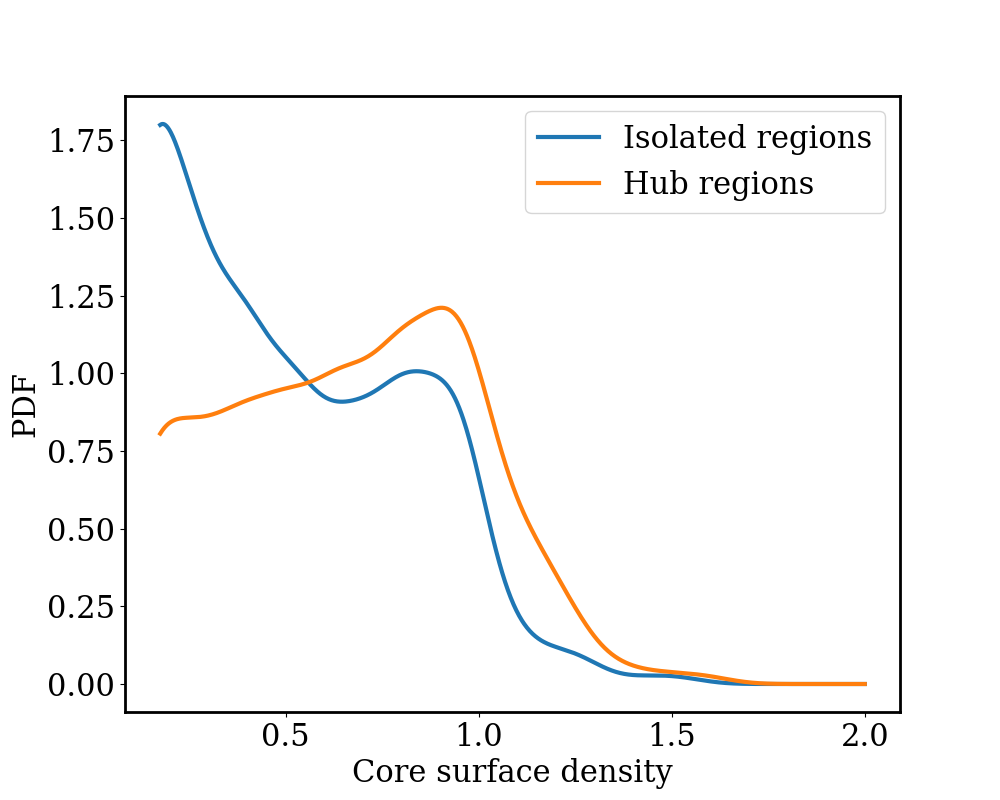}
\caption{A probability density function of the core surface density for all pixels in all 10 simulations belonging to isolated or hub regions. Isolated sub-filaments typically host no core or one core, while hubs of sub-filaments usually host at least one core. The distributions are normalised such that their integral are equal to 1. The bandwidth of both KDEs is 0.027.}
\label{fig::cd_hist}
\end{figure}

Junctions of sub-filaments may also harbour a higher density of cores than single sub-filaments. We thus compare the sub-filament surface density to the core surface density. The core surface density is constructed by convolving the core locations with a two-dimensional Gaussian kernel with a standard deviation of 4 pixels, the same standard deviation as that used for the sub-filament surface density. We divide the map into hub sub-filament regions (the sub-filament surface density is greater than 1.9), and isolated sub-filament regions (the sub-filament surface density lies between 0.9 and 1.1). Figure \ref{fig::cd_hist} shows a KDE of the core surface density in isolated and hub regions, constructed using only map pixels with a core surface density above 0.1, or around 8 pixels away from a core location. Isolated sub-filament regions are skewed towards low core surface densities, peaking at 0.1, and also showing a smaller peak at 1. However, hub sub-filament regions have their maximum likelihood at a core density of around 1. Above a core surface density of 1, the hub sub-filament regions show a clear excess of pixels over the isolated sub-filament regions. Thus, hub sub-filaments show a more clustered form of core formation than that seen in isolated sub-filaments. A two sample Kolmogorov-Smirnov test returns a distance statistic of 0.22 and a corresponding p-value of $10^{-69}$, therefore this difference is statistically significant.

We are thus able to build a coherent scenario of how sub-filaments impact core formation in filaments. Filaments first fragment into sub-filaments due to their internal turbulence, which is both inherited from the large scale flows (i.e. the gas which initially formed the filament) and maintained and driven by sustained accretion. Regions where numerous sub-filaments meet and interact form. We term these hub-sub-filament systems. Such hub-sub-filament systems can be seen in Orion using N$_2$H$^+$ \citep{Hac18}. Isolated sub-filaments fragment into a single core, or a small chain of cores, independent of each other, while hub sub-filaments lead to a small ensemble of cores to form in the hub. Cores formed in such hubs do not only appear more clustered but also more massive than those that form on isolated sub-filaments. This is analogous to core formation in much larger hub-filament systems \citep{Mye09,Per13,Per14,Wil18}.

\begin{figure*}
\includegraphics[width=0.95\linewidth]{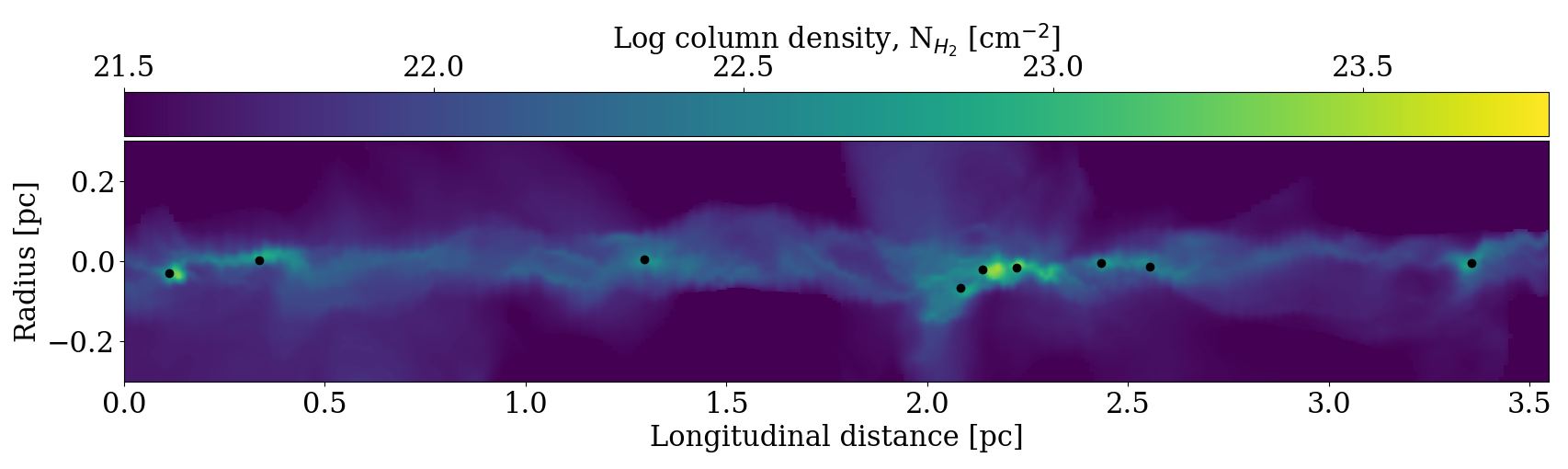}
\caption{The filament from \textsc{SIM01} is shown straightened. Overlaid are black dots showing the core centre of mass positions mapped to this longitude-radius space. This is defined such that spine follows the $r=0$ axis.}
\label{fig::straight_example}
\end{figure*}

\subsection{Fragmentation spacing}\label{SSEC:FRAGSPACE}%

While sub-filaments are intimately linked to core formation in fibrous filaments, there may exist imprints of the fragmentation process of the main filament which is apparent in the core spacing. The widths of the filaments is between 0.1-0.2 pc and are in agreement with the observational results of \citet{Arz11} and \citet{Arz19}. According to equilibrium models we may expect a characteristic core spacing of 0.4-0.8 pc. However, as shown by \citet{Cla16} and \citet{Cla17}, the characteristic fragmentation length-scale is unconnected to the filament width for non-equilibrium filaments. In this section we use the fragmentation analysis tool package \textsc{FragMent} to detect the presence of characteristic fragmentation length-scales.

Before applying the \textsc{FragMent} tool one must use the spines of the main filaments to straighten the filaments. As detailed in \citet{Cla19}, the process of straightening a filament is required to reduce the dimensionality of the problem and remove the complexity of a filament's curvature. The function used to straighten the filaments is \textsc{Straighten$\_$filament$\_$weight}, included in the \textsc{FragMent} package. The parameters used are: \textsc{n$\_$pix}=90, \textsc{max$\_$dist}=30, \textsc{order}=10 and \textsc{h$\_$length}=0.5. The function to map the core positions to the straightened filaments is \textsc{Map$\_$cores}, also included in \textsc{FragMent}, and the value of the parameter \textsc{order} is 10. 

Figure \ref{fig::straight_example} shows the filament from \textsc{SIM01} which has been straightened and the cores, which have been identified using dendrograms, mapped onto this straightened filament. Here one can see that the cores now lie on, or close to, the longitudinal axis of the filament and their positions can be used to investigate the presence of characteristic fragmentation length-scales.

\begin{figure*}
\includegraphics[width=0.48\linewidth]{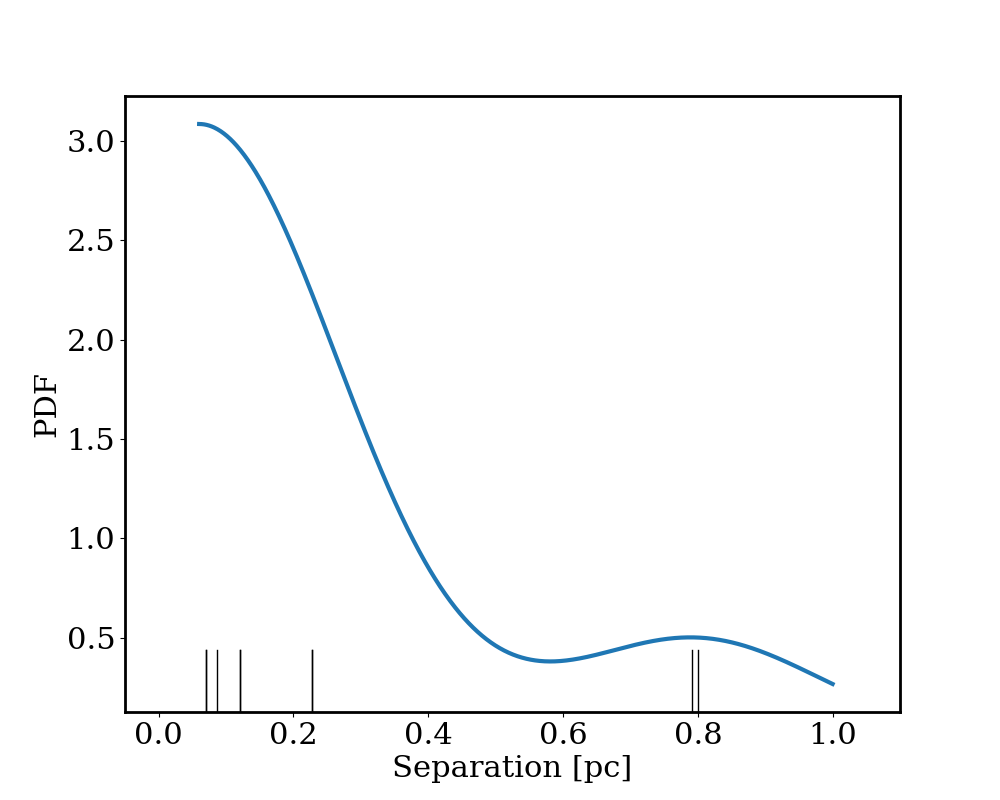}
\includegraphics[width=0.48\linewidth]{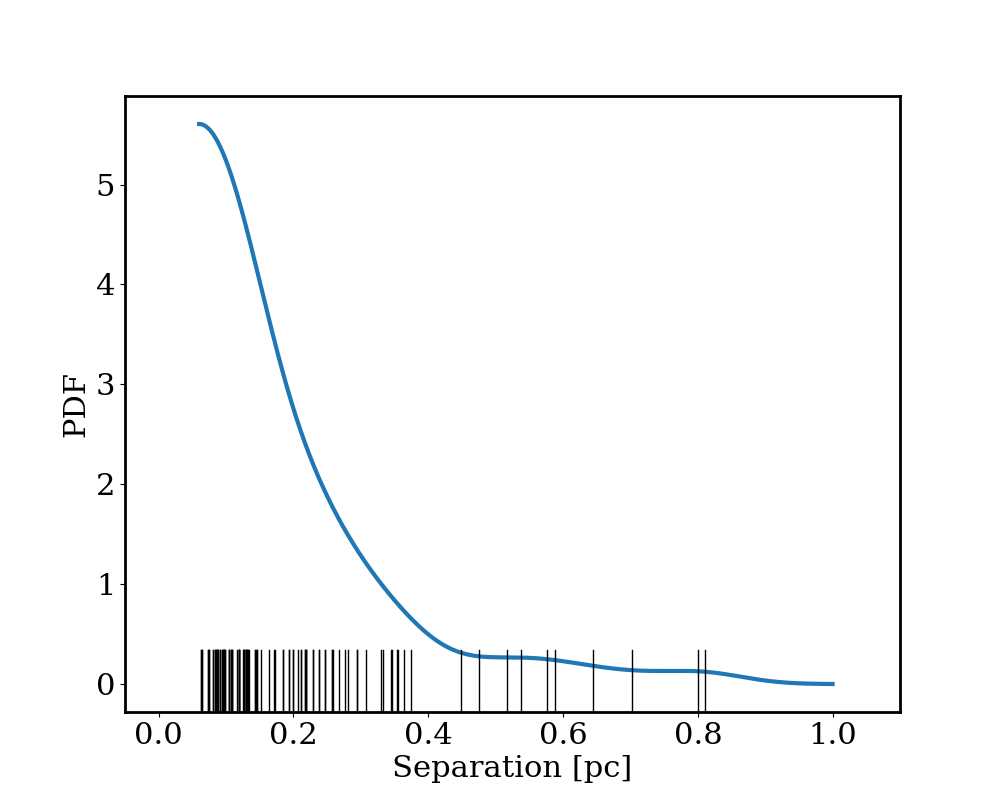}
\caption{(Left) A KDE showing the distribution of nearest neighbour separations taken from \textsc{SIM01}. The KDE bandwidth is 0.15 pc. (Right) A KDE showing the distribution of nearest neighbour separations from all 10 simulations. The KDE bandwidth is 0.03 pc. The small vertical black lines indicate the location of each data point.}
\label{fig::NNS}
\end{figure*}

\begin{table*}
\centering
\begin{tabular}{@{}*6l@{}}
\hline\hline
Nearest neighbour separations & (section \ref{SSSEC::NNS}) & & & \\ \hline
\textsc{SIM}                  & Number of cores & Median [pc] & Mean [pc]   & Minimum $p$-values & Rejection rate\\ \hline
01                            & 9               & 0.12 (0.14) & 0.28 (0.28) & 0.236 (MS)         & 0/4\\
02                            & 12              & 0.12 (0.14) & 0.16 (0.12) & 0.126 (AD)         & 0/4\\
03                            & 12              & 0.13 (0.18) & 0.21 (0.17) & 0.387 (MI)         & 0/4\\
04                            & 11              & 0.14 (0.18) & 0.18 (0.10) & 0.315 (MI)         & 0/4\\
05                            & 12              & 0.12 (0.14) & 0.16 (0.08) & 0.229 (MI)         & 0/4\\
06                            & 13              & 0.13 (0.08) & 0.14 (0.06) & 0.014 (AD)         & 1/4\\
07                            & 16              & 0.11 (0.05) & 0.14 (0.09) & 0.034 (AD)         & 1/4\\
08                            & 9               & 0.25 (0.29) & 0.33 (0.20) & 0.147 (KS)         & 0/4\\
09                            & 13              & 0.10 (0.10) & 0.16 (0.13) & 0.088 (KS)         & 0/4\\
10                            & 9               & 0.21 (0.05) & 0.23 (0.13) & 0.292 (KS)         & 0/4\\ \hline
All simulations               & 116             & 0.12 (0.14) & 0.19 (0.15) & 0.065 (AD)         & 0/4\\ \hline
\end{tabular}
\centering
\caption{A summary of the results from section \ref{SSSEC::NNS} showing the average and widths of the nearest neighbour separation distributions for each simulation. The values in brackets in columns 3 and 4 are the interquartile range and standard deviation, respectively. The letters in brackets in column 5 denote the null hypothesis test which corresponds to the minimum $p$-value: MS is the mean-standard deviation test, MI is the median-interquartile range test, KS is the Kolmogorov-Smirnov test and AD is the Anderson-Darling test. The rejection rate shown in column 6 shows the number of null hypothesis tests which return a $p$-value below 0.05 and allow us to reject the null.}
\label{tab::NNS}
\end{table*}

For the null hypothesis tests included in \textsc{FragMent} one needs a boundary box within which one places randomly-placed cores, defined by the minimum and maximum radius and length. These are different for each simulation. The minimum and maximum values for the length are taken from the spine. The minimum and maximum values for the radius are taken as the minimum and maximum radial locations of the cores, multiplied by 1.1. Typically the spine is roughly 3-4 pc long and the radial width is $<$ 0.1 pc, resulting in a box with an aspect ratio of $\gtrsim$ 30-40; sufficiently high to confidently investigate the mainly longitudinal spacings. 

We use all four null hypothesis tests included in \textsc{FragMent}, the Kolmogorov-Smirnov and Anderson-Darling tests, and the two variants of the average-width test where either the mean and standard deviation or the median and interquartile range of the data distribution is compared to the average and width measurements resulting from from the null hypothesis. All null hypothesis tests are run using 100,000 realisations and with a minimum separation of 0.06 pc. A value of 0.06 pc is taken as it corresponds to two times the minimum effective core diameter from the dendrogram parameter, \textsc{min$\_$npix} = 9. The exact value is arbitrary, with the only constraint being that it is smaller than the smallest separation, though it should be informed by the beam size (in real observations) and the minimum allowed size used by the core finding algorithm. As a consequence of this minimum separation value, when constructing the KDEs in the following two subsections we use a reflective boundary condition at 0.06 pc to avoid the presence of artificial peaks. 

The separation statistics and the results of the null hypothesis tests resulting from the following sections are summarised in tables \ref{tab::NNS} and \ref{tab::MST}. We take 0.05 as the threshold $p$-value for rejecting the null hypothesis.

\subsubsection{Nearest neighbour separation}\label{SSSEC::NNS}%

\begin{figure*}
\includegraphics[width=0.48\linewidth]{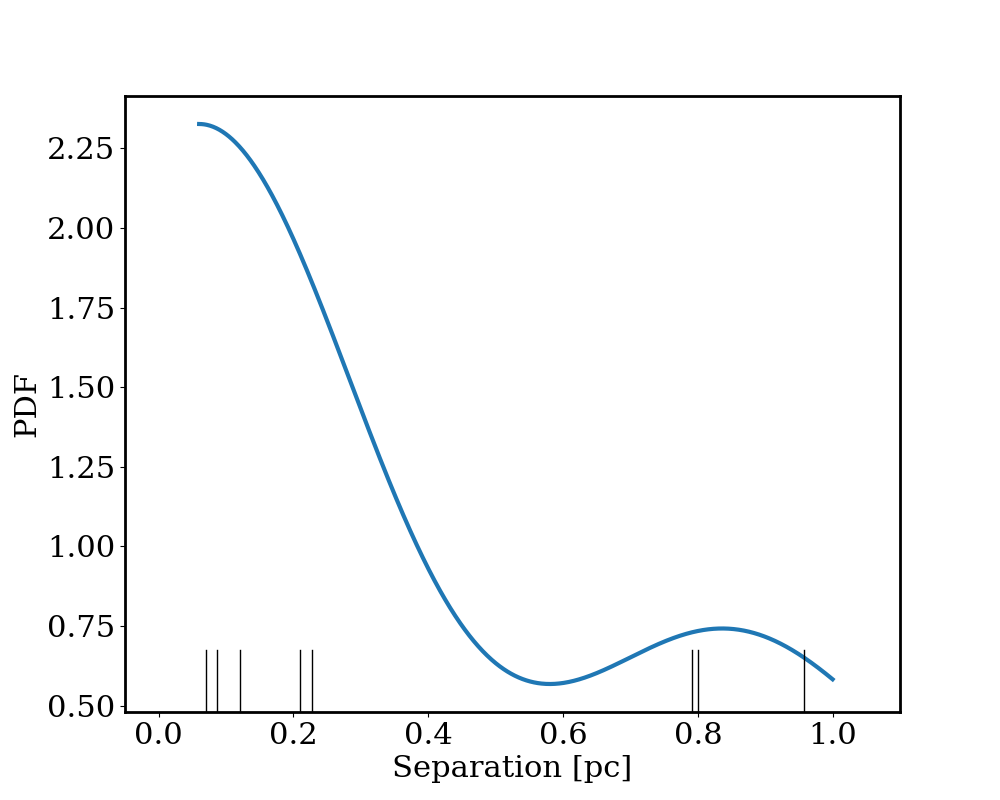}
\includegraphics[width=0.48\linewidth]{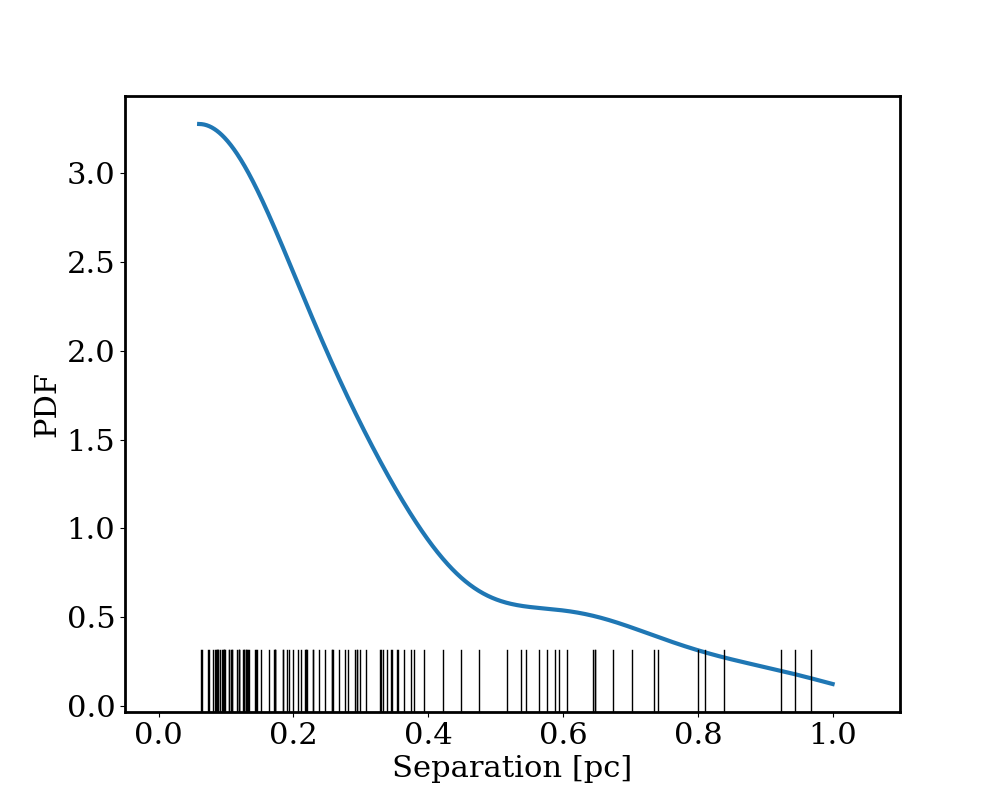}
\caption{(Left) A KDE showing the distribution of minimum spanning tree edge lengths taken from \textsc{SIM01}. The KDE bandwidth is 0.20 pc. (Right) A KDE showing the distribution of minimum spanning tree edge lengths from all 10 simulations. The KDE bandwidth is 0.05 pc. The small vertical black lines indicate the location of each data point.}
\label{fig::MST}
\end{figure*}

\begin{table*}
\centering
\begin{tabular}{@{}*6l@{}}
\hline\hline
Minimum spanning tree edge lengths & (section \ref{SSSEC::MST}) & & & \\ \hline
\textsc{SIM}                  & Number of cores & Median [pc] & Mean [pc]   & Minimum $p$-values & Rejection rate\\ \hline
01                            & 9               & 0.22 (0.68) & 0.41 (0.35) & 0.007 (MI)         & 1/4\\
02                            & 12              & 0.20 (0.25) & 0.25 (0.16) & 0.490 (MI)         & 0/4\\
03                            & 12              & 0.16 (0.30) & 0.27 (0.20) & 0.285 (MI)         & 0/4\\
04                            & 11              & 0.26 (0.24) & 0.28 (0.19) & 0.359 (MI)         & 0/4\\
05                            & 12              & 0.21 (0.16) & 0.23 (0.17) & 0.692 (MS)         & 0/4\\
06                            & 13              & 0.17 (0.20) & 0.27 (0.22) & 0.226 (AD)         & 0/4\\
07                            & 16              & 0.12 (0.22) & 0.21 (0.15) & 0.138 (MI)         & 0/4\\
08                            & 9               & 0.41 (0.36) & 0.42 (0.22) & 0.242 (MI)         & 0/4\\
09                            & 13              & 0.18 (0.23) & 0.27 (0.23) & 0.419 (MS)         & 0/4\\
10                            & 9               & 0.26 (0.39) & 0.40 (0.27) & 0.399 (MI)         & 0/4\\ \hline
All simulations               & 116             & 0.21 (0.25) & 0.29 (0.23) & 0.292 (AD)         & 0/4\\ \hline
\end{tabular}
\centering
\caption{A summary of the results from section \ref{SSSEC::MST} showing the average and widths of the minimum spanning tree edge length distributions for each simulation. The values in brackets in columns 3 and 4 are the interquartile range and standard deviation, respectively. The letters in brackets in column 5 denote the null hypothesis test which corresponds to the minimum $p$-value: MS is the mean-standard deviation test, MI is the median-interquartile range test, KS is the Kolmogorov-Smirnov test and AD is the Anderson-Darling test.The rejection rate shown in column 6 shows the number of null hypothesis tests which return a $p$-value below 0.05 and allow us to reject the null.}
\label{tab::MST}
\end{table*}

Figure \ref{fig::NNS} shows the nearest neighbour separation distribution for \textsc{SIM01}, and the separation distribution taken from combining the results from all 10 simulations. \textsc{SIM01} contains 9 cores and the total sample contains 116. For \textsc{SIM01}, the distribution is dominated by a peak at small separations, close to the separation limit of 0.06 pc, but has a secondary peak at higher separations, $\sim 0.8$ pc. The total distribution is similar but the peak at small separations is considerably stronger and exhibits a tail that extends to higher separations. The small separation peak shows the prevalence of core/clump fragmentation within the filament. The width of both distributions is comparable to the average separation (see table \ref{tab::NNS}) suggesting a lack of a characteristic fragmentation length-scale.

For \textsc{SIM01}, all four null hypothesis tests return $p$-values greater than 0.05; therefore, we can not reject the null hypothesis that the cores are randomly placed. We repeat this analysis for the other 9 simulations and find that in only two of the simulations the null can be rejected (\textsc{SIM06} and \textsc{SIM07}). However, these can only be rejected using one of the tests, the Anderson-Darling test. The other $p$-values for these simulations range from 0.06 to 0.98. We therefore consider it only a tentative rejection. Thus, studying each filament separately leads to no statistically significant and robust fragmentation spacing being detected using the nearest neighbour approach. This may be surprising considering the peaked distributions seen in figure \ref{fig::NNS}, but the null hypothesis distribution is itself peaked. This is due to the fact one has a fixed length filament with a certain number of cores lying in it, and so introduces a length-scale to the problem. This highlights the need for the null hypothesis test and not to rely on a peaked distribution. The results for each simulation are summarised in table \ref{tab::NNS}.

We therefore test the significance of the total separation distribution. As each filament has a unique boundary box due to their varying lengths and the radial distances of the cores, sampling the separation distribution resulting from the null hypothesis is slightly more involved. For each simulation we randomly place the same number of cores as detected in the data in that simulation's boundary box. We then produce the nearest neighbour separation distribution resulting from these randomly placed cores. We do this for each of the 10 simulations and combine the distributions to produce the total nearest neighbour separation distribution from the 116 randomly placed cores. This is treated as 1 realisation of the null hypothesis' distribution and is repeated 100,000 times to produce a well-sampled null hypothesis distribution. None of the four resulting $p$-values are below 0.05. Despite the large sample size the null can not be rejected and the core fragmentation is indistinguishable from randomly placed cores. Note that by combining the results from all 10 simulations we make the implicit assumption that all 10 of the filaments share a common characteristic fragmentation length-scale. This is not necessarily the case and complicates the combination of data from multiple filaments. We thus emphasise that the most important quantity is the number of cores per filament, as pointed out in \citet{Cla19}.

\subsubsection{Minimum spanning tree}\label{SSSEC::MST}%

Figure \ref{fig::MST} shows the minimum spanning tree edge length distribution for \textsc{SIM01}, and the distribution taken from combining the results from all 10 simulations. The distribution resulting from \textsc{SIM01} looks similar to the results from the nearest neighbour separation method, a dominant peak at $\sim 0.1$ pc with a small secondary peak at $\sim 0.8$ pc. The same is true for the total distribution; however, the shoulder feature at $\sim$ 0.6 pc suggests that the distribution is bimodal, a narrow distribution around $\sim$ 0.2 pc and a wider one around 0.6 pc. The width of both distributions are comparable to their average values, making a characteristic fragmentation length-scale unlikely.  

For \textsc{SIM01} the average-width null hypothesis test returns $p$-values of 0.007 using the median-interquartile range, but the other three tests return $p$-values much greater than 0.05. We therefore consider this a tentative rejection. When this analysis is repeated for the other nine simulations we find that the null can not be rejected in any of the simulations. This is similar to the result from the nearest neighbour approach. The results for each simulation are summarised in table \ref{tab::MST}.

For the total distribution, all four null hypothesis tests return $p$-values greater than 0.05 and we can not reject the null hypothesis. 

\subsubsection{Two-point correlation function}\label{SSSEC:2point}%

\begin{figure}
\includegraphics[width=0.95\linewidth]{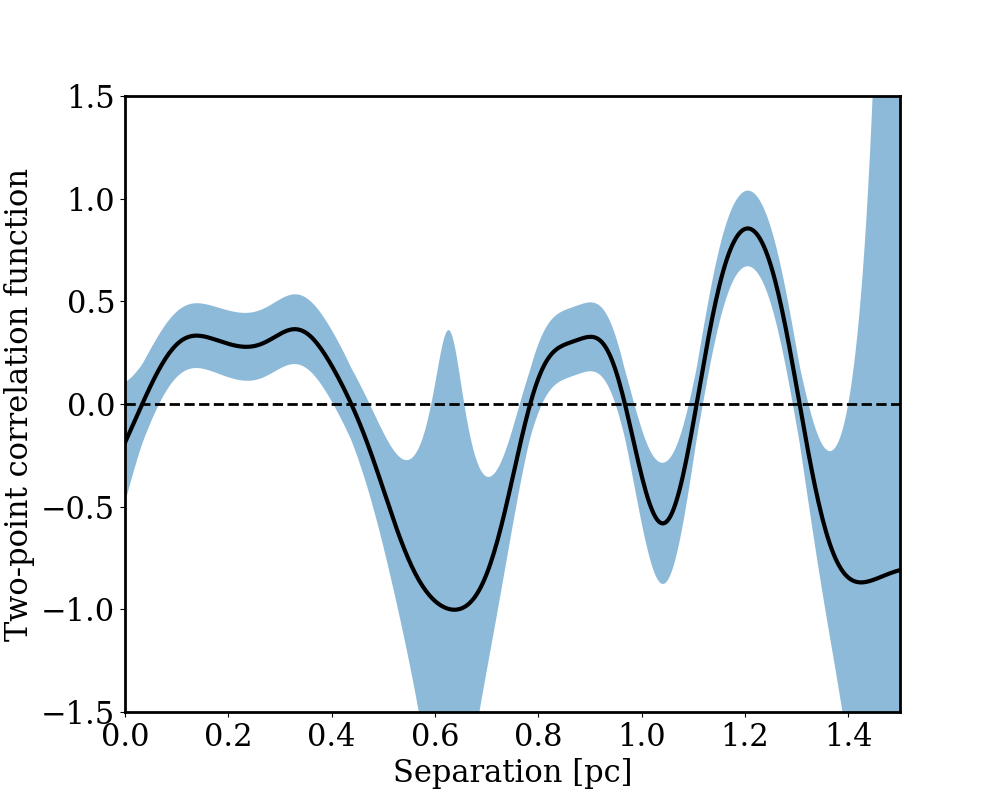}
\caption{The two-point correlation function resulting from \textsc{SIM01}. The blue shaded region shows the 1$\sigma$ errors as each point due to Poisson noise. The horizontal dashed line at $y=0$ is to help guide the reader.}
\label{fig::2point}
\end{figure}

Figure \ref{fig::2point} shows the two-point correlation function for \textsc{SIM01}. The shaded area denotes the 1$\sigma$ error as described in section \ref{SEC:TECH}. There exists an extended feature above zero between 0.1 and 0.4 pc, which correspond to the short spacings seen in the minimum spanning tree and nearest neighbour methods. However, it is clear that this feature is approximately a 2$\sigma$ feature and so is only marginally significant. The next peaks occur at approximately 0.9 and 1.2 pc. The first peak is similarly significant, $\sim 2\sigma$, while the second is considerably more, lying around $5\sigma$ above zero. The peak at $\sim$0.9 pc corresponds to the larger scale separations seen in the minimum spanning tree and nearest neighbour methods. However, the peak at $\sim$1.2 pc does not. \citet{Cla19} show that two-tier fragmentation produces peaks in the two-point correlation function which correspond to superpositions of the underlying characteristic fragmentation length-scales, i.e. if there are two length-scales at $x_1$ and $x_2$, the two-point correlation function may present peaks at $x_2 - x_1$ and at $x_2 + x_1$. As the difference between the peaks at $\sim$ 0.9 pc and 1.2 pc, 0.3 pc, corresponds to the feature between 0.1 and 0.4 pc, we suggest that this is tentative evidence of two-tier fragmentation. However, it is unclear why the signal from the superposition is much stronger than the signal at the two underlying characteristic fragmentation length-scales, or why, at the expected location of the $x_2 - x_1$ superposition, there is a deficit of separations (though statistically insignificant due to the error size). As noted by \citet{Cla19}, signatures of two-tier fragmentation in the two-point correlation function are complex and typically larger number statistics are needed.

As there is no way to combine the results of two-point correlation functions from multiple different data sets we present the individual two-point correlation function plots from the other 9 simulations in appendix \ref{app::2point}. In general, there is no strong signature in any of the two-point correlation functions and no coherent picture can be formed. This is similar to the results from using the nearest neighbour separation and minimum spanning tree methods.

\subsubsection{Lack of a characteristic fragmentation length-scale}\label{SSSEC:LACK}%

There exists no strong signature for the existence of a characteristic fragmentation length-scale in any of the 10 simulations, or the combination of all 10 simulations. The results from the nearest neighbour and minimum spanning tree methods are unable to pass the null hypothesis test, and the results from the two-point correlation function are tentative at best. While this could be due to the small number of cores, \citet{Cla19} show that only a few cores are typically needed to detect single-tier fragmentation, $N\sim10$; a condition which is satisfied here. There may exist an underlying two-tier, or more complex, fragmentation pattern but we are unable to detect it as $N>20$ is typically needed. We therefore can not rule out the null hypothesis, that there exists no characteristic fragmentation length-scale in fibrous filaments. This could be due to the fact that the fragmentation into cores proceeds via sub-filaments, rather than directly from the main filament. \citet{Cla17} show that the formation of sub-filaments and hubs is linked to the turbulent gas motions within the filament, and so has no preferred length-scale. This intermediate fragmentation step erases the expected characteristic fragmentation length-scales that has been seen in previous works of fibre-less filaments \citep{InuMiy92,InuMiy97,FisMar12,Cla16,Cla17}. We thus suggest that filaments containing fibres/sub-filaments should lack characteristic fragmentation length-scales, while those without such substructure are more likely to exhibit a fragmentation length-scale due to the dominance of gravity.  

This result may appear in contradiction to recent observational work \citep{Jac10,Bus13,Lu14,Beu15,Hen16,Tei16,Kai17,Lad20,Zha20}. These works used a variety of methods to investigate the presence of a characteristic fragmentation length-scale; some used null hypothesis tests and others did not, some relied on clustering in the core separations and others used more advanced techniques such as the two-point correlation function and the N$^{th}$ nearest neighbour method. As shown here, and in more detail in \citet{Cla19}, a null hypothesis test is necessary whichever method one uses; clustering or peaks in the separation distribution does not allow one to draw strong conclusions. Moreover, \citet{Cla19} show that certain techniques such as the N$^{th}$ nearest neighbour method are fairly insensitive to the presence of characteristics length-scales and may even produce spurious signatures of such a length-scale when the distribution of cores is random. They also show that the construction of the random core positions for the two-point correlation function must be handled carefully if one wishes to investigate filament fragmentation, i.e. the random cores should be placed on or very close to the filament spine rather than the entire map. Thus, in light of this recent work one should be cautious about the presence of characteristic fragmentation length-scales when these techniques have been used. A re-examining of the data with the techniques presented here and in \citet{Cla19} is encouraged.

The lack of a characteristic fragmentation length-scale does not change when we consider the simulations from a different viewing angle (see appendix \ref{app::proj}).

\subsection{End-dominated collapse}\label{SSEC:END}%

\begin{figure}
\includegraphics[width=0.98\linewidth]{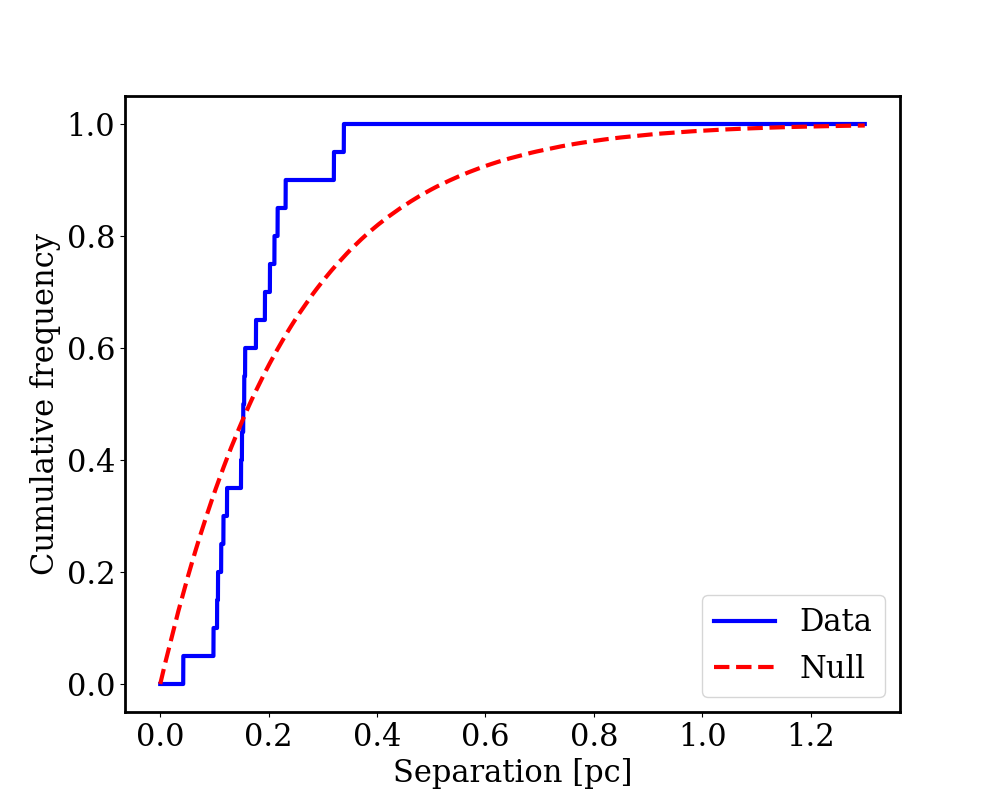}
\caption{The cumulative probability function of the distance between end cores and the spine beginning and end. The solid blue line shows the distribution derived from the simulations, the dashed red line shows the distribution assuming the null hypothesis.}
\label{fig::end_dist}
\end{figure}

While there may not exist a characteristic fragmentation length-scale due to filament fragmentation, there may exist another signature of core formation in filaments. Due to non-linear terms in the gravitational acceleration becoming important at the ends of a filament, global collapse proceeds via an \textit{end-dominated} mode \citep{Bas83}. This means that the ends of the filament are accelerated to greater speeds than the filament interior, sweeping up gas and becoming dense \citep[see][ and references therein for more details on this phenomenon]{Cla15}. Thus, dense and massive cores should preferentially form close to the filament ends. Observations support that this end-dominated collapse may occur in isolated filaments and induce star formation \citep{Zer13,Beu15,Kai16,Dew19,Yuan20,Bha20,Liu20}.

The ends of the filaments are defined as the two longitudinal parts at either end of the filament where its column density drops to the background value. This is non-obvious in the case of a non-ideal filament without a sharp boundary at either end. Here we use the first and last pixel of the spine to denote the ends of the filament as the main spine is defined by a column density cut, as seen in section \ref{SSEC:MAINFIL}.  

We first investigate if cores do preferentially form at the ends of filaments. We do this by measuring the distance between the start of the spine and the first core, and the distance between the last core and the end of the spine. We call these cores the \textit{end} cores, and the remainder of the cores \textit{interior} cores. We perform a null hypothesis test where the null hypothesis is that cores are placed randomly and thus there is no preferential formation of cores near the filament ends. We use the same 100,000 random realisations as those used in the preceding section. 

Figure \ref{fig::end_dist} shows the cumulative probability function for the simulations and resulting from the null hypothesis test. The two distributions have very similar medians, 0.15 and 0.17 pc for the distribution resulting from the data and the null hypothesis, respectively. However, the null hypothesis distribution extends to considerably larger separations while the distribution from the data is steeper. This is evident in the larger interquartile range of the null hypothesis distribution, 0.26 pc, compared to the data distribution, 0.09 pc. We apply a two sample Kolmogorov-Smirnov test and Anderson-Darling test to evaluate if the distribution from the data is distinct from that resulting from the randomly placed cores. For the KS test, the test statistic is large, 0.29, with a corresponding $p$-value of 0.06, and the AD test also returns a large test statistic, 2.40, and a $p$-value of 0.03. We thus tentatively reject the null hypothesis.

\begin{figure*}
\includegraphics[width=0.48\linewidth]{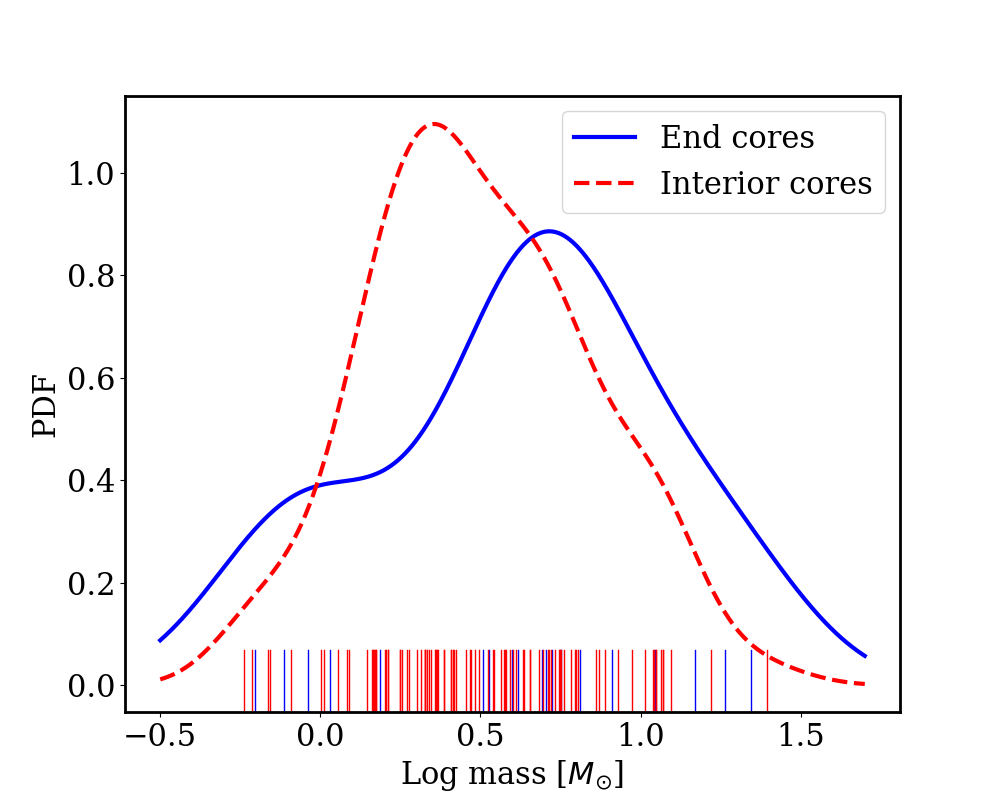}
\includegraphics[width=0.48\linewidth]{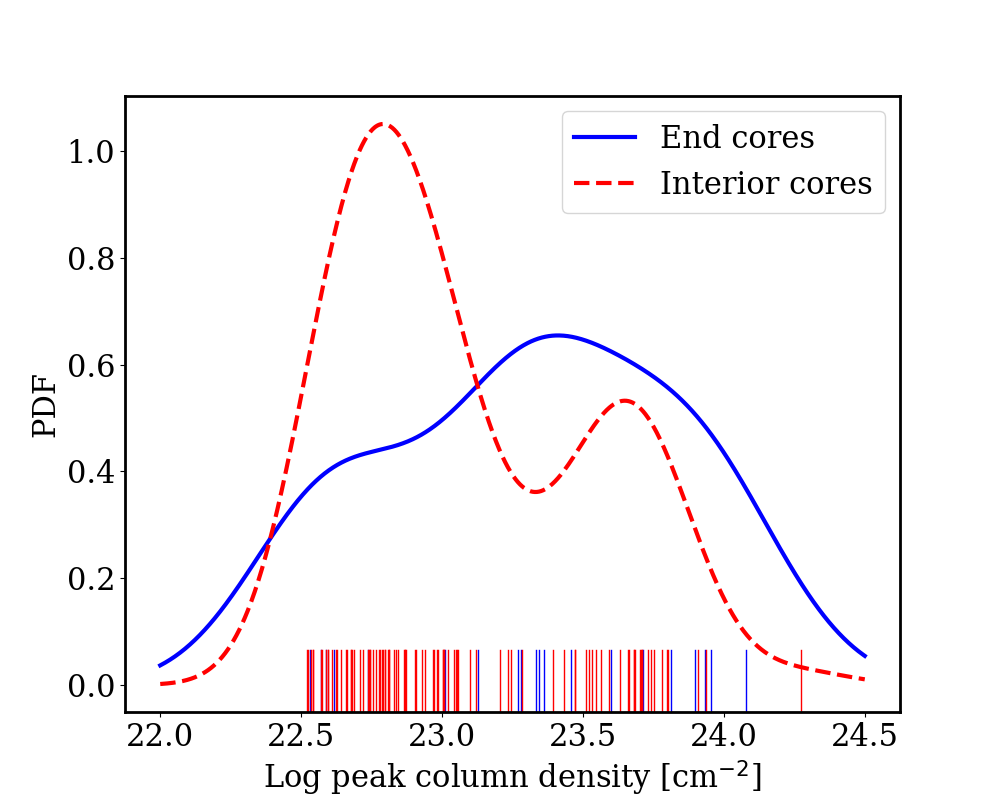}
\caption{(Left) The mass distribution of end and interior cores from all ten simulations in solid blue and dashed red lines, respectively. The KDE bandwidth for the end cores is 0.18, and 0.09 for the interior cores. (Right) The peak column density distribution of end and interior cores from all ten simulations. There are large differences between the interior and end core distributions. The KDE bandwidth for the end cores is 0.21, and 0.11 for the interior cores. The distributions are normalised such that their integrals are equal to 1.}
\label{fig::end_clump}
\end{figure*}

We next ask if these end cores are more massive than the interior cores. The left panel of figure \ref{fig::end_clump} shows kernel density estimators of the mass distribution of end cores and interior cores. One can see that the mass distribution for end cores is shifted to higher masses compared to the interior cores. The mean and standard deviation of the mass distributions are: 6.4 M$\Su$ and 5.8 M$\Su$ for the end cores, and 4.3 M$\Su$ and 3.9 M$\Su$ for the interior cores. A KS test returns a test statistic of 0.29 with a corresponding $p$-value of 0.10, and the AD test returns a test statistic of 1.27 with a $p$-value of 0.10. The Mann-Whitney U test returns similar $p$-value of 0.08. We are therefore unable to reject the null hypothesis that the masses of end cores and interior cores are sampled from the same underlying distribution, but the test statistics are relatively large and more simulations could help reject the null.

We also compare the peak column density of end cores to that of interior cores. The right panel of figure \ref{fig::end_clump} shows kernel density estimators of the peak column density distribution of end cores and interior cores. One can see that the distribution for end cores is peaked at slightly higher column densities compared to the bimodal interior core distribution. The mean and standard deviation of the column density distributions are: 3.5 $\times 10^{23}$ cm$^{-2}$ and  3.4 $\times 10^{23}$ cm$^{-2}$ for the end cores, and 2.1 $\times 10^{23}$ cm$^{-2}$ and  2.6 $\times 10^{23}$ cm$^{-2}$ for the interior cores. A KS test returns a test statistic of 0.36 with a corresponding $p$-value of 0.02, and the AD test returns a test statistic of 2.39 with a $p$-value of 0.03. The Mann-Whitney U test also returns a similarly small $p$-value of 0.04. We therefore reject the null hypothesis and can say that the peak column density of end cores is statistically higher than those of interior cores.

We conclude that there is a tentative signature of end-dominated collapse, leading to the formation of cores close to the filament ends which are slightly more massive and dense than cores located in the filament's interior. An increase in number statistics would likely increase this signature due to the large test statistic values returned by the KS and AD tests.

\section{Conclusions}\label{SEC:CON}%

Our numerical study presents an intriguing scenario of how sub-filaments impact fragmentation. First, filaments fragment into sub-filaments due to their internal turbulence, likely driven by accretion from the surrounding medium. Regions where a number of sub-filaments join and meet form; we term these regions hub-sub-filament systems due to their similarity to the parsec-scale hub-filament systems commonly seen in molecular clouds \citep[e.g. SDC13 and MonR2,][]{Wil18,Trev19}. These hubs fragment into small ensembles of clustered cores. Away from these hubs, sub-filaments fragment into isolated single cores, or a small chain of cores. Cores formed in the hubs are on average more massive than those formed in the isolated sub-filaments, and also show a wider mass distribution. This is also reminiscent of the parsec-scale hub-filament systems, leading to the conclusion that the combination of turbulence and gravity leads to similar patterns of fragmentation on multiple scales.  

The fact that filaments first fragment into sub-filaments which then proceed to form cores erases any evidence of the expected characteristic fragmentation length-scale of filament fragmentation. This fits well with observations which have been unable to find robust evidence for quasi-periodically spaced cores, and is a significant departure from previous fragmentation models. We thus expect that filaments containing sub-filaments to show no clear sign of characteristic fragmentation length-scales, while those filaments without sub-filaments are more likely to. 

End-dominated collapse leads to the preferential formation of cores close to a filament's ends. These cores are slightly more massive and dense than cores located in the interior due to gravitational focusing. However, this signature is weak and requires good number statistics for a robust detection. 

As this work used an idealised cylindrical colliding flow set-up (e.g. without magnetic fields) to investigate the basic underlying physics of filament fragmentation, large-scale molecular cloud simulations would be a valuable follow up, allowing the addition of environmental complexities. Such a study (Ganguly et al. in prep.) will be carried out within the SILCC-zoom project \citep{Wal15,Sei17}, which simulates the formation of molecular clouds from the galactic, multi-phase interstellar medium. 

\section{Acknowledgments}\label{SEC:ACK}%
We thank the referee for their comments which have helped to improve the quality of the paper. SDC and SW acknowledges support from the ERC starting grant No. 679852 `RADFEEDBACK'. GMW acknowledges support from the UK's Science and Technology Facilities Council under grant number ST/R000905/1. SW thank the DFG for funding through the Collaborative Research Center (SFB956) on the `Conditions and impact of star formation' (sub-project C5).

\section{Data availability}\label{SEC:ACK}%
The data underlying this article will be shared on reasonable request to the corresponding author. The principle analysis tool of this work, \textsc{FragMent}, is made freely available at https://github.com/SeamusClarke/FragMent.

\bibliographystyle{mn2e}
\bibliography{ref}

\appendix

\newpage

\section{The two-point correlation function results for \textsc{SIM02} - \textsc{SIM010} }\label{app::2point}%

\begin{figure*}
\includegraphics[width=0.9\linewidth]{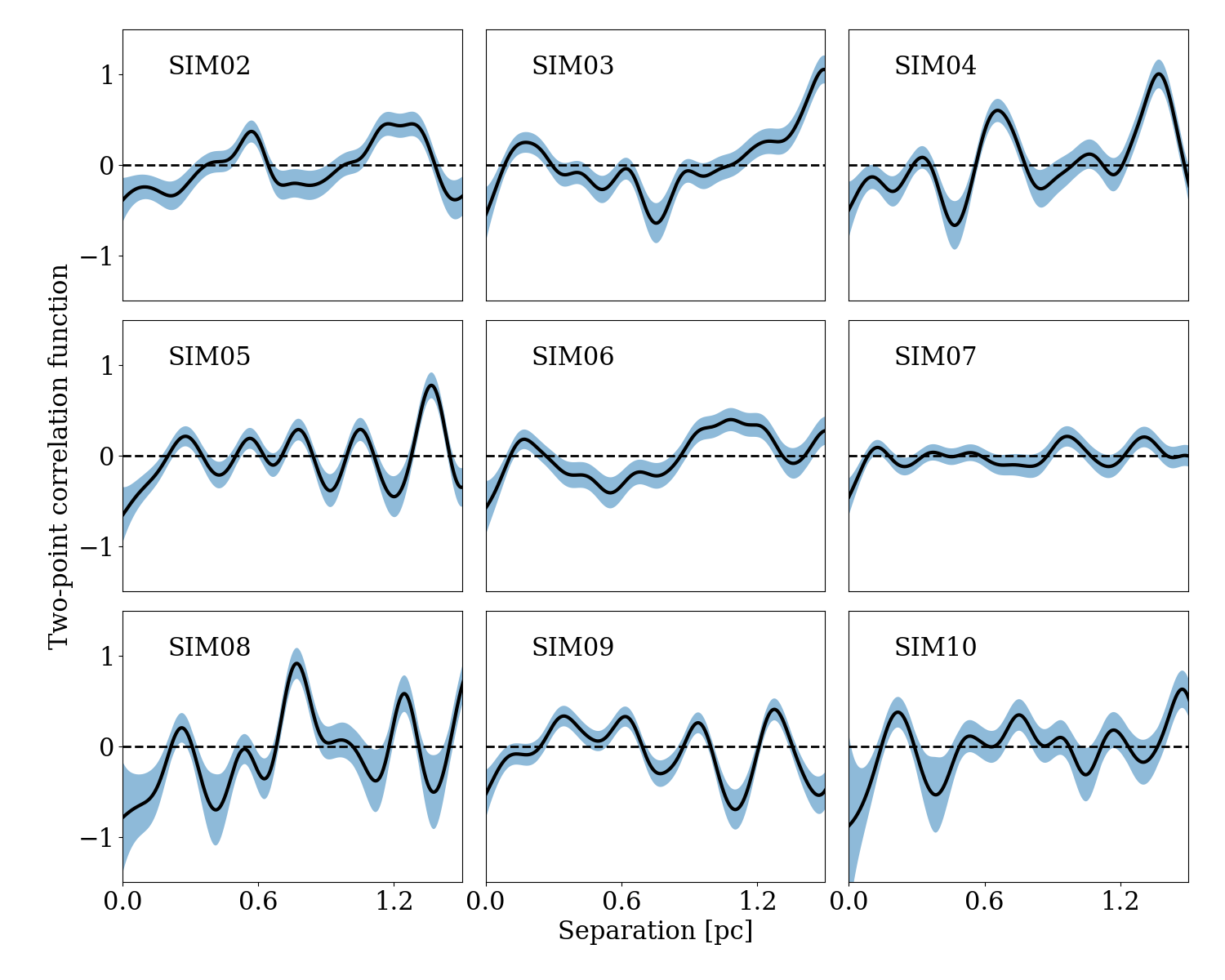}
\caption{The two-point correlation function resulting from \textsc{SIM02} to \textsc{SIM10}. The blue shaded region shows the 1$\sigma$ errors at each point due to Poisson noise. The horizontal dashed line at $y=0$ is to help guide the reader.}
\label{fig::2pointappendix}
\end{figure*}

The individual two-point correlation functions resulting from the nine other simulations, \textsc{SIM02} - \textsc{SIM10}, are shown in figure \ref{fig::2pointappendix}. The results from each are summarised in table \ref{tab::2point}. There is no clear sign of any characteristic fragmentation length-scale.

\begin{table*}
\centering
\begin{tabular}{@{}*3l@{}}
\hline\hline
& Two-point correlation function & (section \ref{SSSEC:2point}) \\ \hline
\textsc{SIM}                   & Inference & Comment\\ \hline
01                             & Tentative two-tier fragmentation    & Possible features at $\sim$0.2 and $\sim$0.9 pc which produce a superposition at $\sim$1.2 pc.\\
02                             & Tentative single-tier fragmentation & Feature at $\sim$0.6 pc with possible harmonic at $\sim$1.2 pc.\\
03                             & No clear signal                     & Weak feature at $\sim$0.2 pc, no harmonics. Peak at 1.5 pc also apparent.\\
04                             & Tentative single-tier fragmentation & Feature at $\sim$0.65 pc and harmonic at $\sim$1.4 pc.\\
05                             & Single-tier fragmentation           & While individually each peak is weak the harmonics are clear at $\sim$0.25, 0.5, 0.75, 1.0, 1.25 pc.\\
06                             & No clear signal                     & Extended feature between 0.8 and 1.2 pc.\\
07                             & No clear signal                     & Nearly zero everywhere.\\
08                             & Tentative two-tier fragmentation    & Possible features at $\sim$0.2 and 0.7 pc with weak superpositions at $\sim$0.5 and 0.9 pc.\\
09                             & Tentative single-tier fragmentation & Possible feature and harmonics at $\sim$0.3, 0.6, 0.9 and 1.3 pc.\\
10                             & No clear signal                     & Close to zeros everywhere with no clear harmonics.\\ \hline
\end{tabular}
\centering
\caption{A summary of the inferences one may derive from the two-point correlation functions from each simulation.}
\label{tab::2point}
\end{table*}


\begin{figure*}
\includegraphics[width=0.98\linewidth]{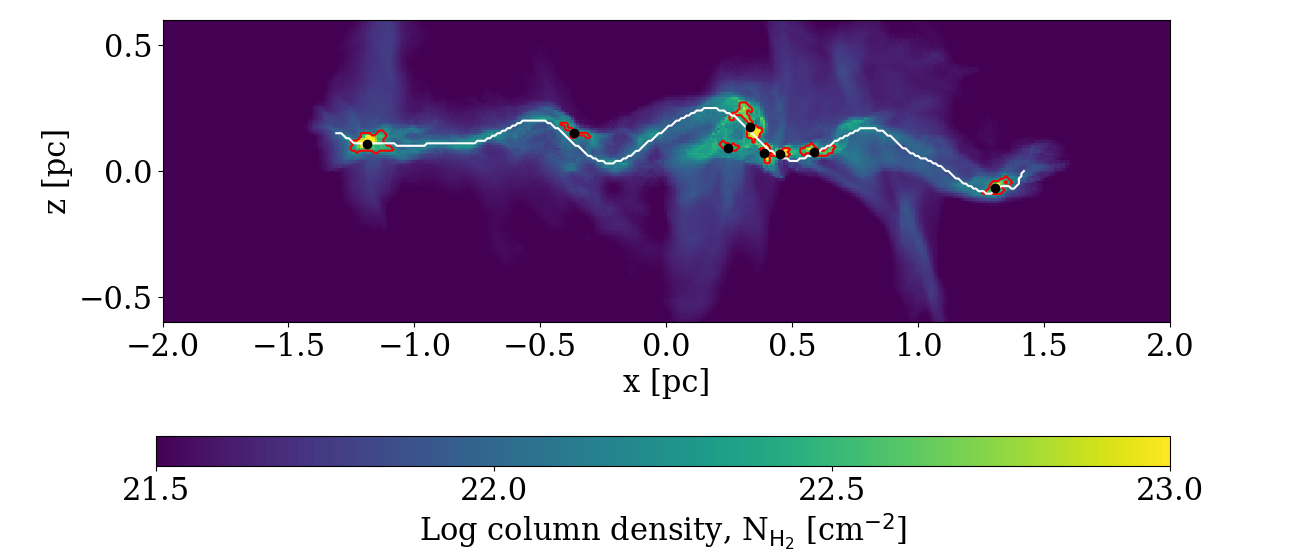}
\caption{The column density plot of \textsc{SIM01} with the appendix viewing angle. The red contours show the core boundaries as found using a dendrogram, with the black dots the column-density weight centre of the cores. The white line shows the main spine of the filament.}
\label{appfig::spine}
\end{figure*}

\begin{figure}
\includegraphics[width=0.98\linewidth]{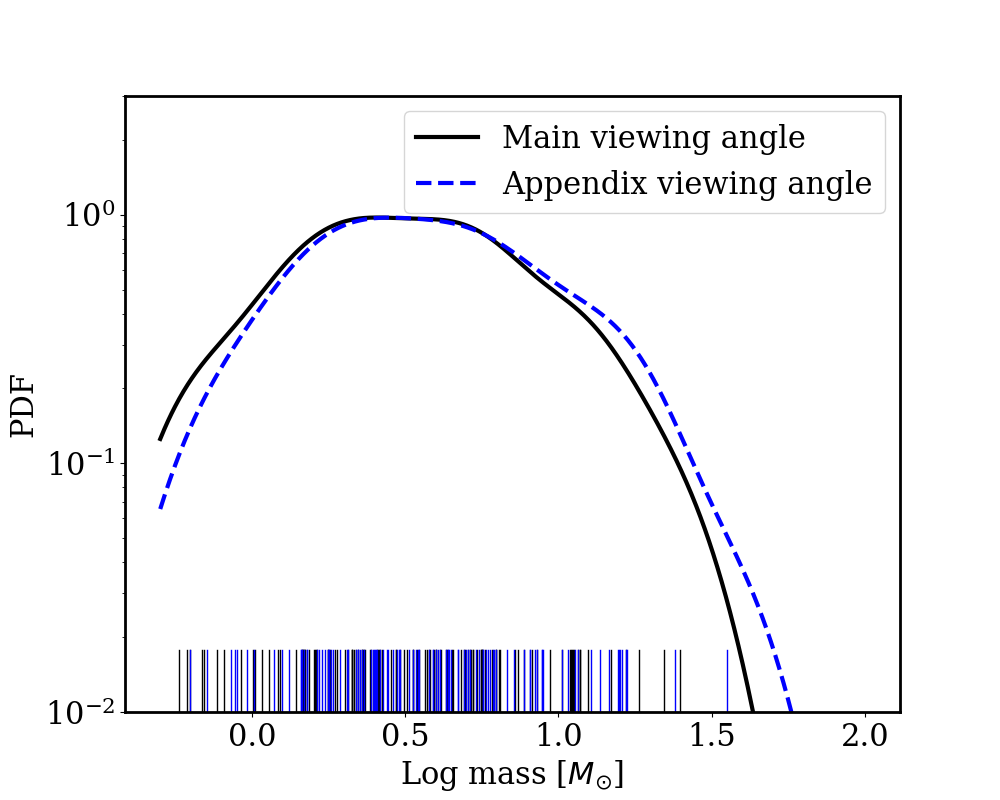}
\caption{The core mass distribution using the cores identified with the main viewing angle (black, solid line), and the appendix viewing angle (blue, dashed line). The small vertical lines show each individual data point.}
\label{appfig::coremass}
\end{figure}

\section{The effect of viewing angle}\label{app::proj}%

As this work has been preformed on column density maps from 3D simulations created by considering a viewing angle aligned with the $z$-axis, we check the effect of projection by considering a viewing angle aligned with the $y$-axis. We call the viewing angle (or projection) aligned with the $z$-axis the \textit{main} viewing angle (or projection) as it is discussed in the main body of the paper, and that aligned with the $y$-axis, the \textit{appendix} viewing angle (or projection).

\subsection{Core and spine identification}\label{app::iden}%

Figure \ref{appfig::spine} shows the column density plot of \textsc{SIM01} from the appendix viewing angle. Overlaid in red are the cores and in white is the main filament spine; both are found using the same techniques and parameters as in sections \ref{SEC:CORE} and \ref{SSEC:MAINFIL}, respectively. From all ten simulations, 105 cores are identified, only slightly fewer than that observed with the main viewing angle. Figure \ref{appfig::coremass} shows the core mass distributions from the two viewing angles, it is clear there are few differences. A two sample KS test returns a small test statistic of 0.08 and a $p$-value of 0.84. Thus, the core masses from the two viewing angles are statistically indistinguishable. 

\subsection{Characteristic fragmentation length-scale}\label{app::frag}%

\subsubsection{Nearest neighbour results}\label{app::NNS}%

\begin{figure*}
\includegraphics[width=0.48\linewidth]{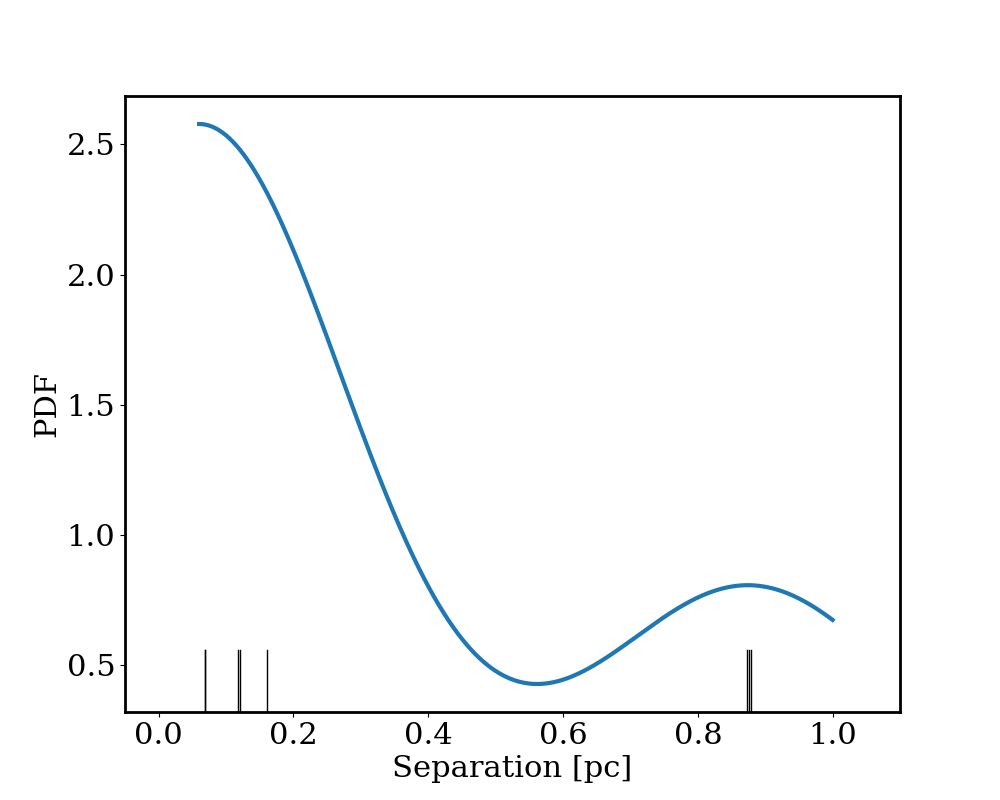}
\includegraphics[width=0.48\linewidth]{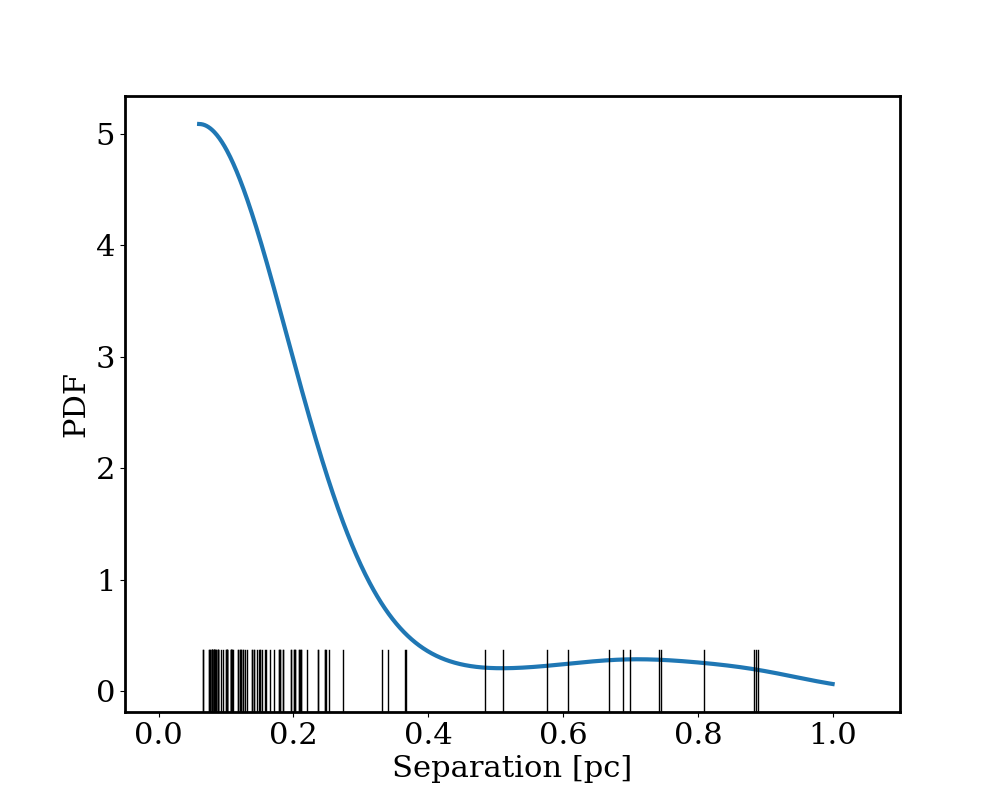}
\caption{(Left) A KDE showing the distribution of nearest neighbour separations taken from \textsc{SIM01}. (Right) A KDE showing the distribution of nearest neighbour separations from all 10 simulations. The small vertical black lines indicate the location of each data point.}
\label{appfig::NNS}
\end{figure*}

Figure \ref{appfig::NNS} shows the nearest neighbour separation distribution from \textsc{SIM01} for the cores identified with the appendix viewing angle, and also the separation distribution when combining results from all 10 simulations. The results are summarised in table \ref{apptab::NNS}. As with the main viewing angle, the results from the nearest neighbour show no strong evidence of a characteristic fragmentation length-scale, and the null hypothesis can not be rejected strongly.

\subsubsection{Minimum spanning tree results}\label{app::MST}%

\begin{figure*}
\includegraphics[width=0.48\linewidth]{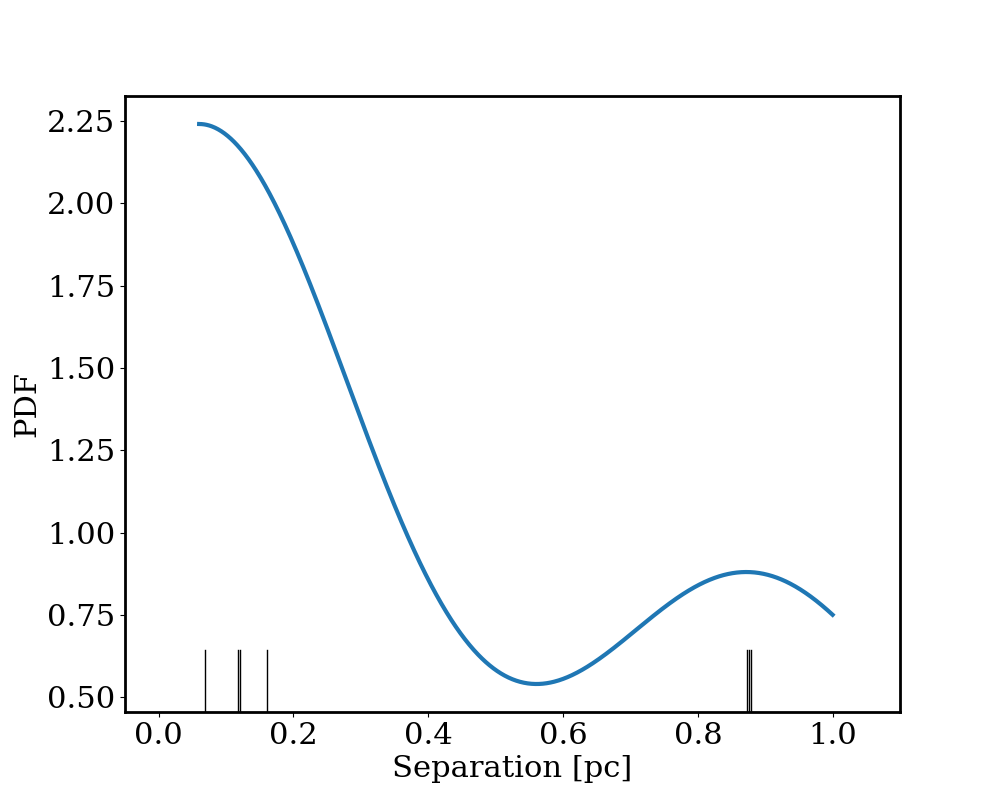}
\includegraphics[width=0.48\linewidth]{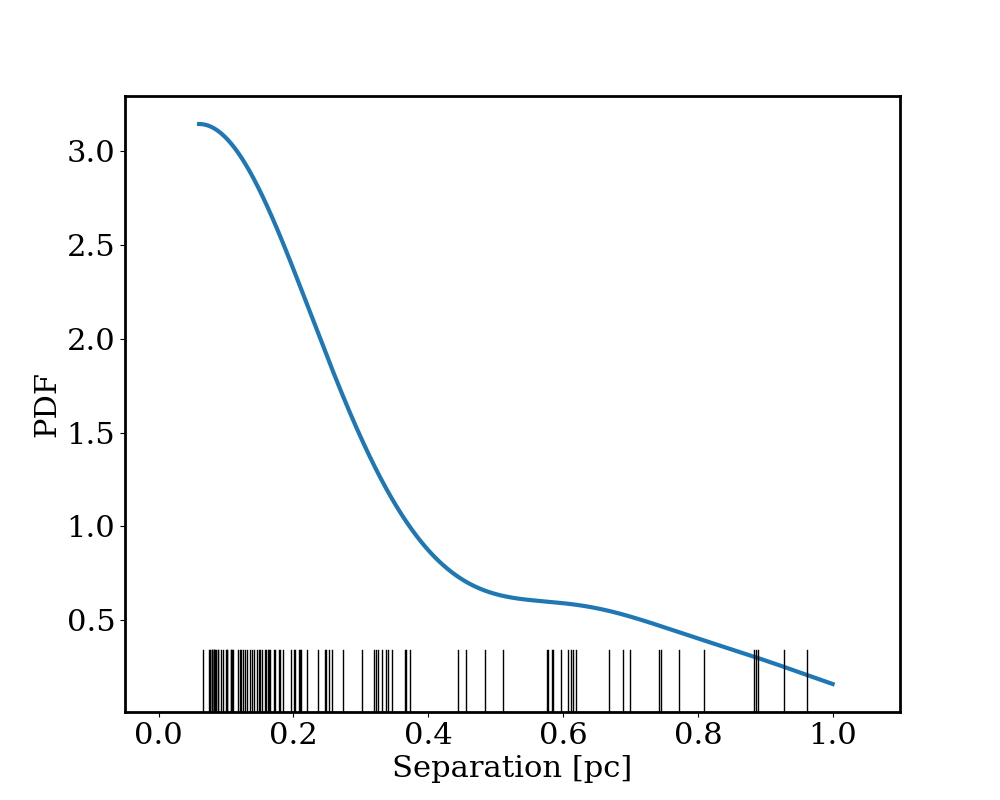}
\caption{(Left) A KDE showing the distribution of minimum spanning tree edge lengths taken from \textsc{SIM01}. (Right) A KDE showing the distribution of minimum spanning tree edge lengths from all 10 simulations. The small vertical black lines indicate the location of each data point.}
\label{appfig::MST}
\end{figure*}

Figure \ref{appfig::MST} shows the minimum spanning tree edge length distribution from \textsc{SIM01} for the cores identified with the appendix viewing angle, and also the edge length distribution when combining results from all 10 simulations. The results are summarised in table \ref{apptab::MST}. As with the main viewing angle, the results from the minimum spanning tree show no strong evidence of a characteristic fragmentation length-scale, and the null hypothesis can not be rejected strongly.

\subsubsection{Two-point correlation function}\label{appB::2point}%

\begin{figure}
\includegraphics[width=0.95\linewidth]{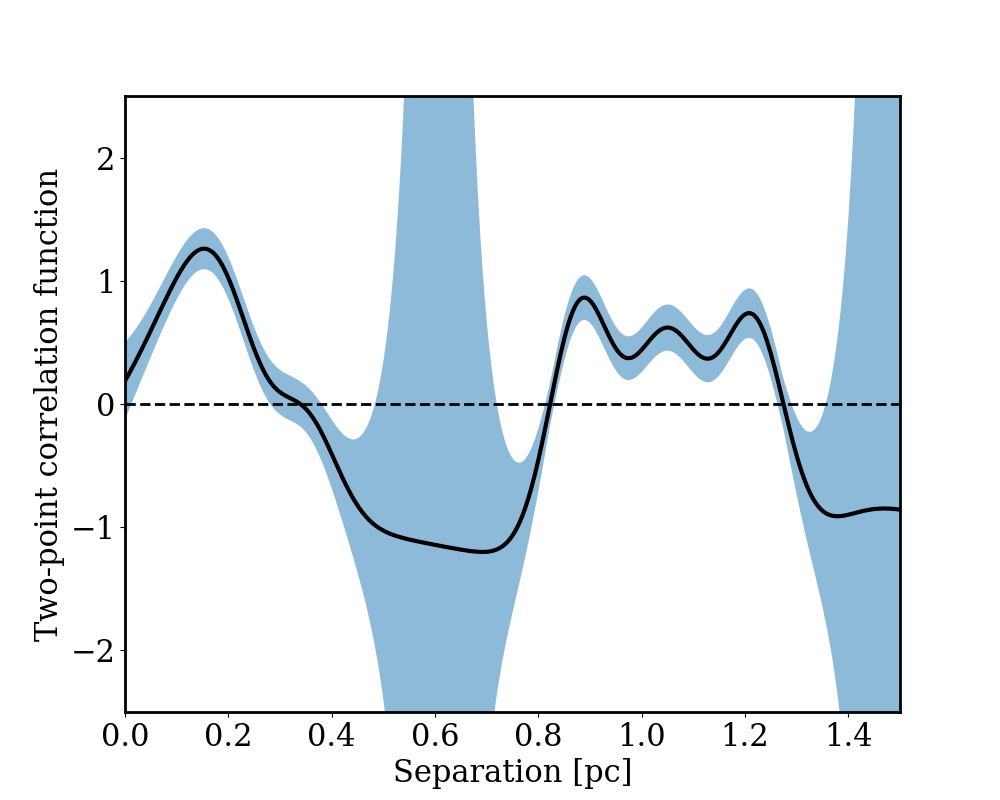}
\caption{The two-point correlation function resulting from \textsc{SIM01}. The blue shaded region shows the 1-sigma errors as each point due to Poisson noise. The horizontal dashed line at $y=0$ is to help guide the reader.}
\label{appfig::2point}
\end{figure}

Figure \ref{appfig::2point} shows the two-point correlation function for \textsc{SIM01} for the cores identified with the appendix viewing angle. Here there are strong signs of two-tier fragmentation, a small-scale peak at $\sim$0.2 pc, and a large scale one at $\sim$1.1 pc, with superpositions at $\sim$0.9 pc and $\sim$1.3 pc. This bimodality is seen in the nearest neighbour and minimum spanning tree results but neither could robustly reject the null hypothesis. 

\begin{figure*}
\includegraphics[width=0.9\linewidth]{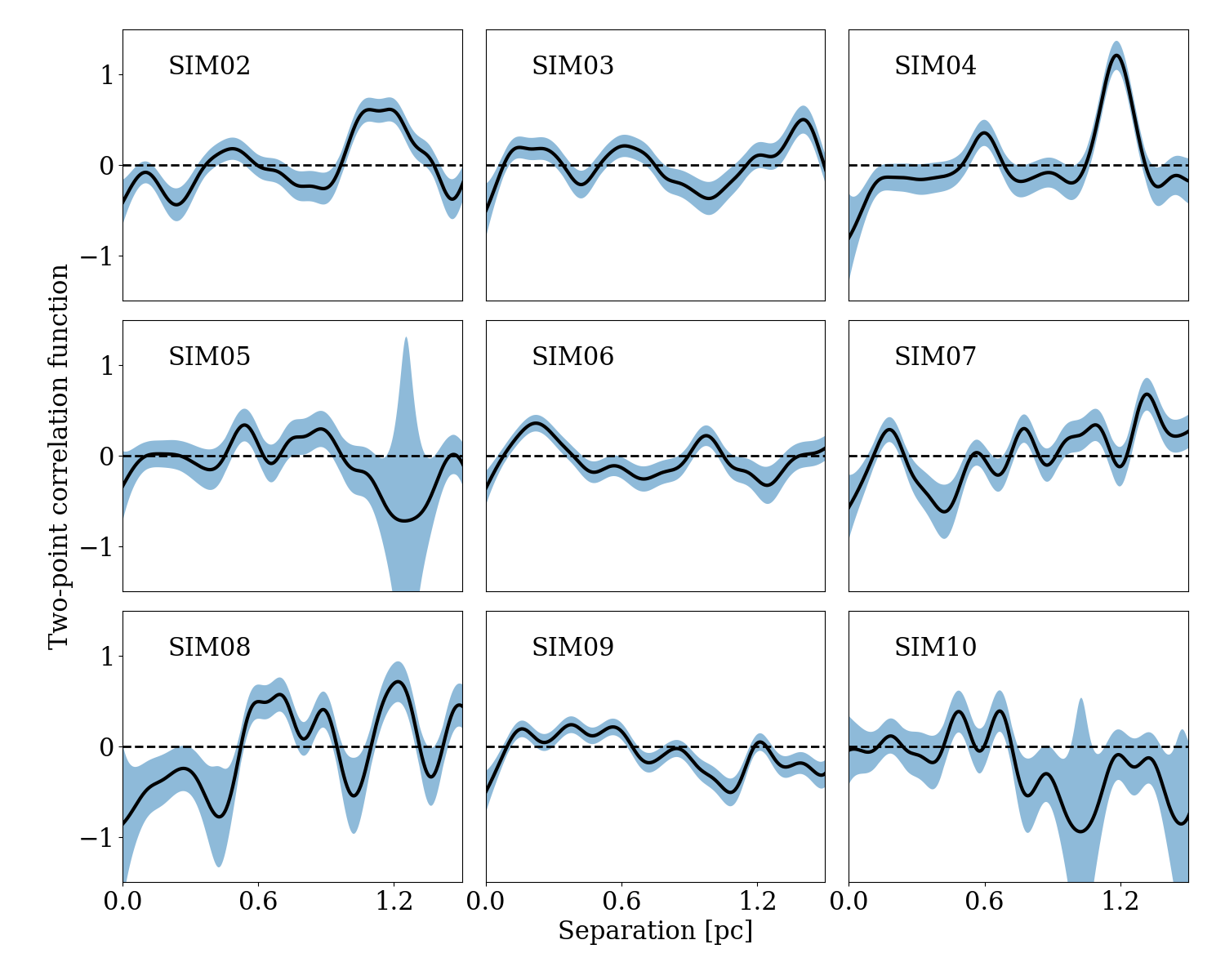}
\caption{The two-point correlation function resulting from \textsc{SIM02} to \textsc{SIM10}. The blue shaded region shows the 1-sigma errors at each point due to Poisson noise. The horizontal dashed line at $y=0$ is to help guide the reader.}
\label{appfig::all2point}
\end{figure*}

Figure \ref{appfig::all2point} shows the two-point correlations functions for \textsc{SIM02} to \textsc{SIM10}. There is no strong evidence for any characteristic length-scale in these nine simulations, other than in \textsc{SIM04} which suggests a possible signal at $\sim$0.6 pc and a harmonic at $\sim$1.2 pc. This is similar to the results presented in the main body of the paper, section \ref{SSSEC:2point}.

\begin{table*}
\centering
\begin{tabular}{@{}*6l@{}}
\hline\hline
Nearest neighbour separations & (section \ref{app::NNS})  & & & \\ \hline
\textsc{SIM}                  & Number of cores & Median [pc] & Mean [pc]   & Minimum $p$-values & Rejection rate\\ \hline
01                            & 8               & 0.14 (0.77) & 0.40 (0.37) & 0.005 (MI)         & 2/4\\
02                            & 12              & 0.11 (0.12) & 0.15 (0.10) & 0.427 (AD)         & 0/4\\
03                            & 11              & 0.13 (0.08) & 0.21 (0.21) & 0.190 (MS)         & 0/4\\
04                            & 10              & 0.13 (0.07) & 0.15 (0.05) & 0.368 (KS)         & 0/4\\
05                            & 9               & 0.15 (0.14) & 0.25 (0.21) & 0.453 (MS)         & 0/4\\
06                            & 15              & 0.09 (0.06) & 0.11 (0.05) & 0.002 (AD)         & 1/4\\
07                            & 10              & 0.19 (0.09) & 0.20 (0.13) & 0.711 (MI)         & 0/4\\
08                            & 8               & 0.27 (0.40) & 0.37 (0.24) & 0.105 (MI)         & 0/4\\
09                            & 15              & 0.10 (0.08) & 0.16 (0.18) & 0.042 (MS)         & 1/4\\
10                            & 7               & 0.17 (0.07) & 0.22 (0.19) & 0.115 (KS)         & 0/4\\ \hline
All simulations               & 105             & 0.14 (0.12) & 0.21 (0.20) & 0.133 (AD)         & 0/4\\ \hline
\end{tabular}
\centering
\caption{A summary of the results from section \ref{app::NNS} showing the average and widths of the nearest neighbour separation distributions for each simulation. The values in brackets in columns 3 and 4 are the interquartile range and standard deviation. The letters in brackets in column 5 denote the null hypothesis test which corresponds to the minimum $p$-value: MS is the mean-standard deviation test, MI is the median-interquartile range test, KS is the Kolmogorov-Smirnov test and AD is the Anderson-Darling test. The rejection rate shown in column 6 shows the number of null hypothesis tests which return a $p$-value below 0.05 and allow us to reject the null.}
\label{apptab::NNS}
\end{table*}

\begin{table*}
\centering
\begin{tabular}{@{}*6l@{}}
\hline\hline
Minimum spanning tree edge lengths & (section \ref{app::MST}) & & & \\ \hline
\textsc{SIM}                  & Number of cores & Median [pc] & Mean [pc]   & Minimum $p$-values & Rejection rate\\ \hline
01                            & 8               & 0.16 (0.75) & 0.44 (0.38) & 0.001 (MI)         & 1/4\\
02                            & 12              & 0.20 (0.25) & 0.25 (0.17) & 0.610 (MS)         & 0/4\\
03                            & 11              & 0.17 (0.40) & 0.28 (0.23) & 0.044 (MI)         & 1/4\\
04                            & 10              & 0.24 (0.18) & 0.28 (0.18) & 0.886 (AD)         & 0/4\\
05                            & 9               & 0.26 (0.40) & 0.38 (0.30) & 0.322 (MI)         & 0/4\\
06                            & 15              & 0.14 (0.12) & 0.22 (0.23) & 0.082 (MI)         & 0/4\\
07                            & 10              & 0.20 (0.35) & 0.32 (0.23) & 0.317 (MI)         & 0/4\\
08                            & 8               & 0.56 (0.37) & 0.46 (0.22) & 0.056 (MI)         & 0/4\\
09                            & 15              & 0.15 (0.21) & 0.23 (0.19) & 0.451 (MI)         & 0/4\\
10                            & 7               & 0.16 (0.76) & 0.44 (0.38) & 0.001 (MI)         & 1/4\\ \hline
All simulations               & 105             & 0.20 (0.35) & 0.32 (0.28) & 0.005 (MS)         & 2/4\\ \hline
\end{tabular}
\centering
\caption{A summary of the results from section \ref{app::MST} showing the average and widths of the minimum spanning tree edge length distributions for each simulation. The values in brackets in columns 3 and 4 are the interquartile range and standard deviation. The letters in brackets in column 5 denote the null hypothesis test which corresponds to the minimum $p$-value: MS is the mean-standard deviation test, MI is the median-interquartile range test, KS is the Kolmogorov-Smirnov test and AD is the Anderson-Darling test.The rejection rate shown in column 6 shows the number of null hypothesis tests which return a $p$-value below 0.05 and allow us to reject the null.}
\label{apptab::MST}
\end{table*}

\label{lastpage}

\end{document}